\documentclass[%
reprint,
nofootinbib,
amsmath,amssymb,
aps,
pre,
]{revtex4-1}
\usepackage{lipsum}
\usepackage{xcolor}
\usepackage{amsmath,amssymb,bbm}
\usepackage{soul}
\usepackage{graphicx}
\usepackage{dcolumn}
\usepackage{bm}
\usepackage[colorlinks=true, citecolor=blue]{hyperref}
\usepackage[mathlines]{lineno}
\usepackage[xcolor,framemethod=TikZ]{mdframed}
\mdfdefinestyle{myStyle}{roundcorner=5pt,backgroundcolor=blue!75!cyan!05,linecolor=blue!40!black,linewidth=1.5pt}
\usepackage{comment}

\usepackage{algorithmicx}
\usepackage[ruled,lined]{algorithm2e}
\usepackage{algpseudocode}

\usepackage{stackengine}

\usepackage{scalerel}
\usepackage{tikz}
\usetikzlibrary{svg.path}
\usepackage{stmaryrd}
\usepackage{enumitem}
\setlist[itemize]{align=parleft,left=0pt..1em}

\usepackage[font=footnotesize,caption=false]{subfig}

\newtheorem{definition}{Definition}

\graphicspath{{./Figures/}}

\begin{document}
	
	\allowdisplaybreaks
	
	\title{Urban traffic analysis and forecasting through shared Koopman eigenmodes}
	
\author{Chuhan Yang}
\affiliation{New York University Abu Dhabi, Saadiyat Island, P.O. Box 129188, Abu Dhabi, U.A.E.}
\author{Fares B. Mehouachi}
\affiliation{New York University Abu Dhabi, Saadiyat Island, P.O. Box 129188, Abu Dhabi, U.A.E.}
\author{M{\'o}nica Men{\'e}ndez}
\affiliation{New York University Abu Dhabi, Saadiyat Island, P.O. Box 129188, Abu Dhabi, U.A.E.}
\author{Saif Eddin Jabari}
\email{sej7@nyu.edu}
\affiliation{New York University Abu Dhabi, Saadiyat Island, P.O. Box 129188, Abu Dhabi, U.A.E.}
	
	
	
\begin{abstract}
Predicting traffic flow in data-scarce cities is challenging due to limited historical data. To address this, we leverage transfer learning by identifying periodic patterns common to data-rich cities using a customized variant of Dynamic Mode Decomposition (DMD): constrained Hankelized DMD (TrHDMD). This method uncovers common eigenmodes (urban heartbeats) in traffic patterns and transfers them to data-scarce cities, significantly enhancing prediction performance. TrHDMD reduces the need for extensive training datasets by utilizing prior knowledge from other cities. By applying Koopman operator theory to multi-city loop detector data, we identify stable, interpretable, and time-invariant traffic modes. Injecting ``urban heartbeats'' into forecasting tasks improves prediction accuracy and has the potential to enhance traffic management strategies for cities with varying data infrastructures. Our work introduces cross-city knowledge transfer via shared Koopman eigenmodes, offering actionable insights and reliable forecasts for data-scarce urban environments.
\end{abstract}
	
\maketitle

\section{Introduction}

Smaller cities face difficulties acquiring data compared to larger cities \citep{wang2020estimating}. This challenge is caused by the substantial maintenance costs associated with the deployment and upkeep of different types of traffic sensors. Comprehensive traffic flow information is also unavailable for most road segments in large road networks. According to ~\citet{lee2023effects}, in the UTD19 dataset \citep{UTD19_link} covering 40 cities (from which 12 cities were later ruled out due to the long aggregation intervals, incomplete measurements, or incomplete loop detector information) that report their loop detector readings, the number of detectors varies from 20 (Stuttgart) to 4736 (London), and the detector density varies from 0.2 per km$^2$ (Stuttgart) to 39.6 per km$^2$ (Frankfurt). With limited information, small cities often lack historical data patterns that traffic operators require for transport system management. This predicament resembles the `cold start' problem \citep{li2022network} commonly encountered in recommender systems research. We aim to explore the possibility of identifying shared characteristics among cities possessing abundant traffic data. Our objective is to understand their unique attributes and discover shared patterns that can be transferred to cities grappling with limited and imbalanced data.

Existing studies have explored the use of multi-source traffic data for various applications, moving beyond traditional loop detectors to incorporate data from vehicle GPS, mobile phones, and wearable devices \citep{lv2016guest,ambuhl2016data,dakic2018use,wang2020estimating,usman2019survey,li2020trajectory}. Other sources, including wireless sensors \citep{ghazal2023internet}, traffic surveillance cameras \citep{fedorov2019traffic}, and radio frequency identification detectors (RFID) \citep{krausz2017radio,liu2019dynamic}, have been integral in amassing multimedia data vital for data-driven management in cities. Nevertheless, these methodologies confront significant challenges. Wireless sensors, for example, suffer from uneven coverage and are constrained by high energy consumption and costs \citep{al2020efficient}. Mobile sensors offer broad spatial coverage but poor temporal coverage, making it difficult to extract temporal patterns \citep{work2008ensemble,huang2018modeling}. Furthermore, RFID sensor reliability diminishes with adverse environmental conditions \citep{cui2019radio}, while traffic cameras necessitate complex video/data analysis \citep{fedorov2019traffic,fredianelli2022traffic}. The heterogeneity of data sources, often maintained by various agencies with differing standards, introduces further complications in data aggregation and integration, exacerbating issues like data insufficiency and inconsistencies due to sensor failures, communication issues, and maintenance \citep{li2022network,yu2023few}. This underscores the continued relevance of knowledge sharing in traffic data collection. In this work, we focus on evaluating our approach only on loop detector measurements across different cities curated by \citet{loder2019understanding}. By doing so, we circumvent the deployment, maintenance, and processing challenges the aforementioned devices encounter, and we enable analysis with limited data. Additionally, loop detectors remain the most prevalent devices providing traffic measurements worldwide \citep{loop2019_Felipe}.

In this work, we focus on city-wide traffic dynamics, which sensor data aggregated at the 3-5 minute level adequately capture and, more importantly, offer important insight for city-wide traffic management. We refer to \citep{rehborn2009traffic,rehborn2020data,kaufmann2018aerial} for works on traffic prediction along single road segments, which aim to capture detailed traffic dynamics, such as synchronized flow and wide-moving jams. At the aggregate (and city-wide) level this article focuses on, several works in the literature have indicated the difficulty of predicting the traffic dynamics \citep{knoop2015traffic,shim2019empirical,saffari2022data,yang2024advanced}. Nevertheless, the importance of this analysis is highlighted by the various city-wide traffic phenomena that have been observed using aggregated data: For instance, clockwise hysteresis loops could have been shown to exist in perfectly symmetric network topologies and uniform demand conditions \citep{gayah2011clockwise,yildirimoglu2015investigating,ambuhl2021disentangling}. Studies have also shown that, in urban networks, the maximum number of congested clusters and the maximum macroscopic fundamental diagram (MFD) flow coincide, indicating a relationship between network percolation and the onset of city-wide congestion \citep{ambuhl2023understanding}. 
Also, perimeter flow patterns have been proven to affect the observed spatial heterogeneity in urban traffic, leading in turn to different congestion patterns and active network bottlenecks \citep{ambuhl2021disentangling}. 
%
%
These observations are important for the deployment of effective control strategies at the city level, which capitalize on the aggregate dynamical features of traffic, e.g., perimeter control \citep{yang2017multi,haddad2017coordinated,li2021robust}, route guidance \citep{yildirimoglu2018hierarchical,hosseinzadeh2023mfd,menelaou2023convexification,chen2024iterative}, parking management \citep{cao2015system,zheng2016modeling,jakob2020dynamic}, pricing \citep{yang2019heterogeneity,loder2022optimal,balzer2022modal,li2022quasi}, and ridesharing \citep{wei2020modeling,beojone2023dynamic,shen2023aggregated}, among others.

In contrast with the existing literature, our work examines city traffic flow as an amalgam of spectral components hereafter called `city heartbeats.' The spectral features include time-invariant modes and associated characteristic variables that encode the dynamical aspects of the time evolution of traffic volumes across the entire city throughout the day, i.e., the growth or decay of the modes. Decomposing the city-wide datasets allows us to discover the dominant features of city traffic using a spectral lexicon. It also enables us to identify fundamental characteristics shared across cities (shared heartbeats) by comparing the frequencies of their modes. We provide a systematic way of understanding patterns that have been known to be reproducible over time \citep{ambuhl2021disentangling} and across cities \citep{ambuhl2023understanding}.

Our methodology employs Koopman operator theory \citep{koopman1932dynamical,koopman1931hamiltonian,rowley2009spectral,budivsic2012applied}, which facilitates the decomposition of nonlinear and non-stationary dynamical systems into time-invariant modes and interpretable dynamical components \citep{mezic2004comparison,mezic2005spectral,mezic2020spectrum,williams2015data,avila2020data}. The theory has been applied across many fields of science and engineering, ranging from dynamical systems and fluid dynamics \citep{schmid2010dynamic,schmid2011applications,brunton2017chaos,nair2020phase} to diverse domains such as wind engineering \citep{li2023koopman}, image processing \citep{bouwmans2016handbook}, game balancing \citep{avila2021game}, quantitative finance \citep{mann2016dynamic}, climate science \citep{chen2023discovering}, multiresolution analysis \citep{dylewsky2019dynamic}, and COVID-19 pandemic prediction \citep{mezic2023koopman}.

On the algorithmic side, our approach is based on dynamic mode decomposition (DMD) \citep{schmid2022dynamic} with Hankelization (HDMD) \citep{arbabi2017ergodic,brunton2017chaos} and frequency-based constrained DMD \citep{krake2022efficient,krake2022constrained}. What sets our strategy apart from Krake et al.'s work \citep{krake2022constrained} is the absence of a human-in-the-loop process. Instead, we employ frequency integration as a validation method to confirm the generalization capability of transferred knowledge extracted across cities. Variants of DMD that address noisy or incomplete observations and memory effects have been proposed in the physics literature \citep{takeishi2017subspace,curtis2021dynamic,anzaki2023dynamic}. 

DMD-based techniques have recently been employed in the urban transport domain \citep{liu2016data,ling2020koopman,wang2023anti,yu2020low,cheng2022real,gu2023deep}. The focus of these works has been local (e.g., a single highway or city) traffic prediction and control. The techniques serve as alternatives to the traditional statistical prediction techniques \citep{van1996combining,jabari2013stochastic,kumar2015short,kumar2017traffic,zheng2018traffic,shahriari2020ensemble}, limited by assumptions of linearity inherent in the methods. For prediction, the absence of hyper-parameter tuning in DMD-based methods and the interpretability of their results gives them an advantage over machine learning methods and deep neural networks \citep{leshem2007traffic,xing2015traffic,ahn2016highway,jabari2019learning,li2020mcshort,benkraouda2020traffic,thodi2021learning,li2022nonlinear,thodi2022incorporating,li2022network,narmadha2023spatio,mendez2023long,yang2023short,wang2023knowledge,wen2023rpconvformer,yang2023short,djenouri2023hybrid,yang2023generalized,thodi2023learning,thodi2024fourier}. We demonstrate the predictive performance of our method as compared to approaches in all of these categories. Our main interest, which has not received attention in the literature, is city-wide spectral features and transferability across cities.

For knowledge transfer across cities, deep learning models are the leading performers; they initially train on extensive data from one city and fine-tune with a smaller dataset from another. However, they face challenges in identifying common features in source and target cities and are sensitive to errors in the data \citep{li2020short,lee2021short,pan2009survey,yan2023identifying,bai2020adaptive,huang2023traffic,kim2019structure}. Alternative methods like multi-class SVMs, kernel regression, and city similarity measures have also been employed in cross-city knowledge transfer \citep{xu2016cross,lin2018transfer,qu2022transfer,wei2016transfer,wang2019ridesharing,guo2018citytransfer}. Our study aligns with zero-shot and few-shot learning, where data scarcity limits traditional training \citep{oreshkin2021meta,dooley2024forecastpfn}. Although conceptually akin to these approaches (namely, \citet{liu2023cross}) in identifying shared traffic patterns across cities, our work diverges in (1) requiring few-shot samples instead of extensive source city data for training and (2) prioritizing the identification and evaluation of common patterns as prior information, rather than training a forecasting model.

\begin{figure*}[!]
\centering
\includegraphics[width=1\textwidth]{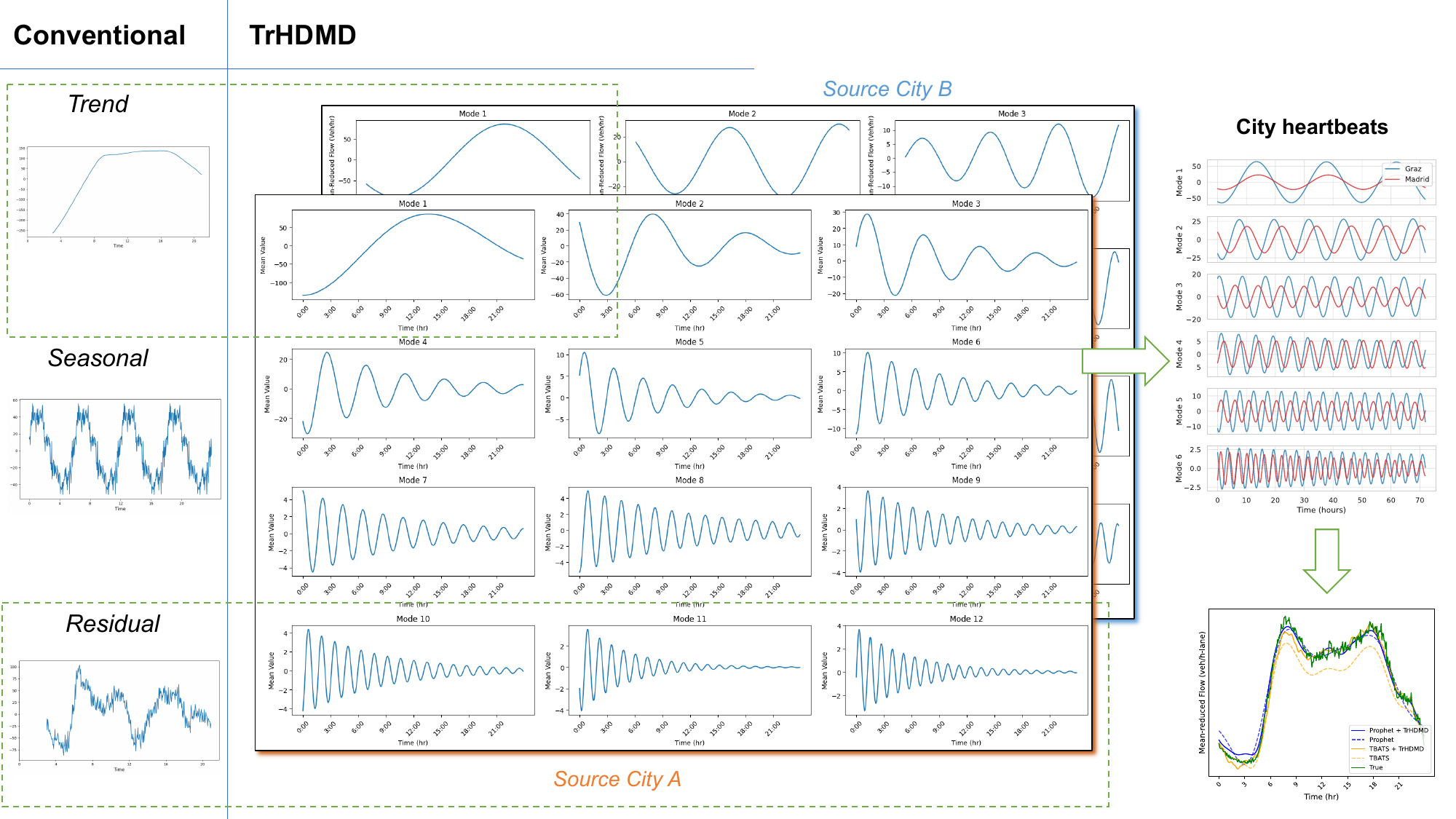}
\caption{Our work examines city traffic flow as an amalgam of spectral components hereafter called `city heartbeats:' We map city dynamics into a space where the nonlinear city traffic dynamics become linear. We perform DMD and break down city traffic into a spectrum of modes characterized by specific frequencies and growth/decay rates, then we identify the shared patterns that could be transferred to downstream tasks in the target city, see Section \ref{sec:methods}. This approach differs from traditional time series decomposition methods, which typically categorize data into trend, seasonal, and residual components. In our framework, all DMD modes can be analogously considered as seasonal components, with modes exhibiting infinite periods akin to trend components, and noisy modes potentially truncated (see Fig.~\ref{fig:eigenvalueDisplay}) serving as residual components. The advantages of our framework can be summarized as follows: (1) Minimal dependency on parameter tuning: We do not rely heavily on precise parameter tuning or extensive training data, the process of delay tuning is described in Section \ref{sec:exp-linearization}. (2) Invariant spectral features: The spectral features extracted are also interpretable in the form of characteristic cycle times instead of high-level features, allowing for clearer explanations of the interface between cities. See Section \ref{sec:exp-eigen-extract} (3) Transferability validation: We evaluate the validity of the transfer by assessing the performance improvement of both Koopman methods and additional time series forecasting methods that require prior information as discussed in Section \ref{sec:exp-transfer-beyond}. }

\label{fig:Tr-HDMD-overview}
\end{figure*}

Fig. \ref{fig:Tr-HDMD-overview} summarizes the proposed framework. The rest of the paper is organized as follows: 
Sec.~\ref{sec:background} provides a background summary of the Koopman operator and dynamic mode decomposition (DMD). Sec.~\ref{sec:methods} presents our methodology, which includes the companion matrix version of DMD and our proposed transfer technique using Henkelized versions of the data matrices, which constitute the embedding we propose for urban traffic. We provide experimental details and results in Sec.~\ref{sec:experiments}. Sec.~\ref{sec:conclusion} concludes the paper.  

\section{Background}\label{sec:background}
\subsection{The Koopman operator}\label{ssec:Koopman}

Consider the discrete-time dynamical system:
\begin{align}
\mathbf{x}_{k+1} &= \mathbf{F}(\mathbf{x}_k),
\end{align}
where $\mathbf{x}_k \in \mathbb{R}^n$ is the system state, $k$ is the index of a discrete timestep, $\mathbf{F}: \mathbb{R}^n \to \mathbb{R}^n$ is a nonlinear vector map describing the state dynamics (state evolution into next time step).

The Koopman operator $\mathcal{K}$ is a linear infinite-dimensional operator that advances functions of the states in time \citep{koopman1931hamiltonian,koopman1932dynamical}: For a scalar-valued function of the state space (an \textit{observable}) $\psi: \mathbb{R}^n \to \mathbb{C}$, the action of $\mathcal{K}$ is
\begin{equation}
(\mathcal{K}\psi)(\mathbf{x}_k) = \psi \circ 
\mathbf{F}(\mathbf{x}_k)  = \psi(\mathbf{x}_{k+1}),
\end{equation}
where $\circ$ denotes the composition of the observation function with the state update map. Essentially, the Koopman operator defines a new dynamical system that \textit{governs the evolution of observables} in discrete time. 

Observables are quantities that are calculated from the state variables. They can represent physical quantities: they can be the states themselves, e.g., in a traffic system, if $\mathbf{x}$ represents the positions of vehicles on a road, $\psi(\mathbf{x})$ can be the first element of $\mathbf{x}$, which is the position of the first vehicle in the system. We will typically have a large number of observables (larger than $n$). 
Observables can also be thought of as coordinate transformations of the state variables.  One technical requirement of the set of observables is that they form a \textit{Koopman-invariant subspace}, that is, when transformed by the Koopman operator, they remain in the same subspace. 

Let $\phi_j: \mathbb{R}^n \to \mathbb{C}$ and $\lambda_j \in \mathbb{C}$ denote an eigenfunction-eigenvalue pair for $\mathcal{K}$, i.e., $\mathcal{K}\phi_j = \lambda_j\phi_j$. Given a dictionary of observables $\{\psi_1,\psi_2,…,\psi_t\}$, we can define the vector-valued function $\Psi:\mathbb{R}^n \to \mathbb{C}^t$ as $\Psi(\mathbf{x}_k) = [\psi_1(\mathbf{x}_k) \cdots \psi_t(\mathbf{x}_k)]^{\top}$, $t \in \{1,...,\infty\}$ is the number of observables. 

To achieve Koopman invariance, we select observables, $\Psi$, the elements of which lie in the span of the eigenfunctions of $\mathcal{K}$, i.e.,
\begin{equation}\label{eqn:KoopmanInvariantObservables}
	\psi_i(\mathbf{x}) = \sum_{j=1}^{\infty} b_{i,j} \phi_j(\mathbf{x}),
\end{equation}
where $b_{i,j} \in \mathbb{C}$ is the $j$th eigenvalue corresponding to the $i$th observable. Consequently,
\begin{equation}\label{eqn:KoopmanOnFObs0}
	\psi_i(\mathbf{x}_k) = (\mathcal{K} \psi_i)(\mathbf{x}_{k-1}) = \sum_{j=1}^{\infty} \lambda_j b_{i,j} \phi_j(\mathbf{x}_{k-1}).
\end{equation}
Equivalently,
\begin{equation}\label{eqn:KoopmanOnFObs1}
	\psi_i(\mathbf{x}_k) = (\mathcal{K}^k \psi_i)(\mathbf{x}_{0}) = \sum_{j=1}^{\infty} \lambda_j^k b_{i,j} \phi_j(\mathbf{x}_{0}).
\end{equation}
Then $\Psi(\mathbf{x})$ can be written as:
\begin{equation}\label{eqn:KoopmanOnFObs}
	\Psi(\mathbf{x}_k) 
	= \sum_{j=1}^{\infty} \lambda_j^k \phi_j(\mathbf{x}_0) \mathbf{b}_j,
\end{equation}
where $\mathbf{b}_j = [b_{1j} ~\cdots~ b_{tj}]^{\top}$. We refer to $\mathbf{b}_j \in \mathbb{C}^t$ as the $j$th \textit{Koopman mode corresponding to the Koopman eigenvalue} $\lambda_j$ associated with the particular (vector) observable $\Psi$. 

\subsection{Dynamic mode decomposition}\label{ssec:dmd}
Dynamic mode decomposition (DMD) is a powerful data-driven algorithm that extracts dynamic features from time series and other dynamical systems. It decomposes a dataset into a spectrum of modes, each associated with a specific frequency and growth/decay rate, thereby revealing the system's intrinsic patterns and coherent structures.

Let $\Psi_k$ denote an observables vector in (discrete) time step $k$ and consider the data matrices
\begin{equation}
	\boldsymbol{\Psi}_{T-1} = \begin{bmatrix}
		\Psi_0 & \cdots & \Psi_{T-1} 
	\end{bmatrix}
	\mbox{ and }~
		\boldsymbol{\Psi}_T = \begin{bmatrix}
		\Psi_1 & \cdots & \Psi_T 
	\end{bmatrix}.
\end{equation}
$\boldsymbol{\Psi}_T$ is time-shifted version of $\boldsymbol{\Psi}_{T-1}$. DMD seeks the eigenvalues and eigenvectors of a high-dimensional matrix $\mathbf{K}$ that relates the two matrices linearly, i.e., $\boldsymbol{\Psi}_{T} \approx \mathbf{K} \boldsymbol{\Psi}_{T-1}$. In essence, $\mathbf{K}$ resembles a finite-dimensional Koopman operator. We define  $\mathbf{K}$ as the solution of an error minimization problem:
\begin{equation}\label{eqn:dmd}
	\mathbf{K} = \underset{\mathbf{A}}{\arg\min}~ \| \boldsymbol{\Psi}_{T} - \mathbf{A} \boldsymbol{\Psi}_{T-1} \|_{\rm F} = \boldsymbol{\Psi}_{T} \boldsymbol{\Psi}_{T-1}^+,
\end{equation}
where $\|\cdot\|_{\rm F}$ is the Frobenius norm and $\boldsymbol{\Psi}_{T-1}^+$ is the Moore-Penrose pseudoinverse of $\boldsymbol{\Psi}_{T-1}$.

Solving for $\mathbf{K}$ directly is computationally prohibitive for the types of high-dimensional systems that the DMD typically solves for. Instead, the spectral features of $\mathbf{K}$ are extracted from a smaller matrix $\widetilde{\mathbf{K}}$ obtained from a similarity transformation (i.e., $\mathbf{K}$ and $\widetilde{\mathbf{K}}$ have the same non-zero eigenvalues). Let $\boldsymbol{\Psi}_{T-1} = \mathbf{U} \boldsymbol{\Sigma} \mathbf{V}^*$ be the (exact or approximate) reduced singular value decomposition (SVD) of $\boldsymbol{\Psi}_{T-1}$. From \eqref{eqn:dmd}, we have that
\begin{equation}\label{eqn:temp}
	\mathbf{K} = \boldsymbol{\Psi}_{T} \mathbf{V} \boldsymbol{\Sigma}^{-1} \mathbf{U}^*.
\end{equation}

Then, utilizing the left singular matrix $\mathbf{U}$ to perform the similarity transformation, from \eqref{eqn:temp} we get the reduced matrix $\widetilde{\mathbf{K}}$ as
\begin{equation}\label{eqn:similarity}
	\widetilde{\mathbf{K}} = \mathbf{U}^* \mathbf{K} \mathbf{U} = \mathbf{U}^* \boldsymbol{\Psi}_T \mathbf{V} \boldsymbol{\Sigma}^{-1}.
\end{equation}
We determine the eigenvectors and eigenvalues of $\widetilde{\mathbf{K}}$, denoted $\widetilde{\mathbf{B}}$ and $\mathsf{diag}(\boldsymbol{\lambda})$, respectively. We recover $\mathbf{B}$ from $\widetilde{\mathbf{B}}$ using
\begin{equation}
	\mathbf{B} = \boldsymbol{\Psi}_T \mathbf{V} \boldsymbol{\Sigma}^{-1} \widetilde{\mathbf{B}} \mathsf{diag}(\boldsymbol{\lambda})^{-1}.
\end{equation}

\section{Methodology}\label{sec:methods}

\subsection{From Koopman to DMD}\label{ssec:KoopmanDmd}
The key to linking the Koopman operator to DMD lies in (i) expressing \eqref{eqn:KoopmanOnFObs} as a matrix decomposition, and (ii) utilizing the relationship between Vandermonde matrices and companion matrices. First, we write  \eqref{eqn:KoopmanOnFObs} as
	\begin{equation} \label{eqn:KoopmanDecomposion1}
		\Psi(\mathbf{x}_k)
		= \begin{bmatrix}
			\vline & \vline &   \\ \mathbf{b}_1 & \mathbf{b}_2 & \cdots  \\ \vline & \vline & 
		\end{bmatrix}
		\begin{bmatrix}
			\phi_1(\mathbf{x}_0) & 0  & \cdots \\
			0 & \phi_2(\mathbf{x}_0)  & \cdots \\
			&  & \ddots
		\end{bmatrix}
		\begin{bmatrix}
			\lambda_1^k \\
			\lambda_2^k \\
			\vdots
		\end{bmatrix}.
	\end{equation}

\begin{mdframed}[style=myStyle]

\smallskip

Consider the $T$-fold evolution of $\Psi$, defined as $\boldsymbol{\Psi}_T \equiv \begin{bmatrix} \Psi(\mathbf{x}_0) & \cdots & \Psi(\mathbf{x}_T) \end{bmatrix}$. We can interpret the rows and columns of $\boldsymbol{\Psi}_T$ as observations (rows) over time (columns):
	\def\tmp{%
		\begin{bmatrix}
			\psi_1(\mathbf{x}_0) & \psi_1(\mathbf{x}_1) & \cdots & \psi_1(\mathbf{x}_T) \\ \psi_2(\mathbf{x}_0) & \psi_2(\mathbf{x}_1) & \cdots & \psi_2(\mathbf{x}_T) \\ \vdots & \vdots & \ddots & \vdots  \\ \psi_t(\mathbf{x}_0) & \psi_t(\mathbf{x}_1) & \cdots & \psi_t(\mathbf{x}_T)
		\end{bmatrix}.
	}%
	\begin{equation}
		\boldsymbol{\Psi}_T =
		\stackMath\def\stackalignment{r}%
		\stackon%
		{\rotatebox[]{90}{\mbox{\scriptsize observables $\approx$ space}}\left\{\tmp\right.}%
		{\overbrace{\phantom{\smash{\tmp\mkern -36mu}}}^{\mbox{\scriptsize time}}\mkern 20mu}%
	\end{equation}
	
From \eqref{eqn:KoopmanDecomposion1}, we can express $\boldsymbol{\Psi}_T$ as the product of three matrices ($\mathbf{B}$, $\boldsymbol{\Phi}$, and $\boldsymbol{\Lambda}$):
	\begin{equation}
		\underbrace{
			\begin{bmatrix}
				\vline & \vline &  \\ \mathbf{b}_1 & \mathbf{b}_2 & \cdots  \\ \vline & \vline & 
			\end{bmatrix}
		}_{\begin{matrix} \mathbf{B} \\ {\mbox{\scriptsize modes}} \end{matrix}}
		\underbrace{
			\begin{bmatrix}
				\phi_1 & 0  & \cdots \\
				0 & \phi_2  & \cdots \\
				&  & \ddots
			\end{bmatrix}
		}_{\begin{matrix} \boldsymbol{\Phi} \\ {\mbox{\scriptsize spectrum/amplitudes}} \end{matrix}}
		\underbrace{
			\begin{bmatrix}
				1 & \lambda_1 & \cdots & \lambda_1^T \\
				1 & \lambda_2 & \cdots & \lambda_2^T \\
				\vdots & \vdots &  \ddots & \vdots
			\end{bmatrix},
		}_{\begin{matrix} \boldsymbol{\Lambda} \\ {\mbox{\scriptsize dynamics}} \end{matrix}}
	\end{equation}
where we have dropped the (trivial) dependence on $\mathbf{x}_0$.

Each row in $\mathbf{B}$ represents an observable (or a spatial dimension if we think of observables as coordinate transformations) and each column of $\boldsymbol{\Lambda}$ represents a time step. The eigenfunctions of $\mathcal{K}$, $\boldsymbol{\Phi}$, play the role of amplitudes in the decomposition.  Notice that the matrix $\boldsymbol{\Lambda}$ is a Vandermonde matrix, which are notoriously ill-conditioned for moderate to large values of $T$. We address this next.

\smallskip

\end{mdframed}

The first step in connecting the Koopman operator to DMD is to enforce a finite-dimensional system. A natural choice for the dimension of the finite-dimensional subspace is $T$. With slight notation abuse, we will use $\mathbf{B}$, $\boldsymbol{\Phi}$, and $\boldsymbol{\Lambda}$ to denote the finite-dimensional versions of their infinite-dimensional counterparts. Then, $\boldsymbol{\Lambda}$ becomes a $T \times T$ square Vandermonde matrix, which diagonalizes a square $T \times T$ companion matrix $\mathbf{C}$. In other words, the matrix
\begin{equation}
	\mathbf{C} \equiv \boldsymbol{\Lambda}^{-1} \boldsymbol{\Phi} \boldsymbol{\Lambda}
\end{equation}
has the structure of a companion matrix, i.e., 
\begin{equation}\label{eqn:Comp}
	\mathbf{C} = \begin{bmatrix}
		0 & \cdots & 0 & c_0 \\
		1 & \cdots & 0 & c_1 \\
		\vdots & \ddots & \vdots & \vdots \\
		0 & \cdots & 1 & c_{T-1}
	\end{bmatrix}
\end{equation}
for $n$ unknown coefficients $c_0,...,c_{T-1}$. Hence,
\begin{equation}
	\boldsymbol{\Psi}_T = \mathbf{B} \boldsymbol{\Phi} \boldsymbol{\Lambda} = \mathbf{B} \boldsymbol{\Lambda} \mathbf{C}.
\end{equation}

While $\mathbf{B}$ and $\boldsymbol{\Lambda}$ are unknown, the action of right multiplying their product by $\mathbf{C}$ is known: the first $T-1$ columns of $\mathbf{C}$ perform column shifts and the last column produces a linear combination. Specifically, the fist $T-1$ columns of  $\mathbf{B} \boldsymbol{\Lambda} \mathbf{C}$ are simply the last $T-1$ columns of $\mathbf{B} \boldsymbol{\Lambda}$ and row $i$ of the last column is $\sum_{j=1}^T c_{j-1}(\mathbf{B} \boldsymbol{\Lambda})_{i,j}$. Thus, \emph{$\mathbf{B} \boldsymbol{\Lambda}$ is just a back-shifted version of $\boldsymbol{\Psi}_T$}, which implies that $\mathbf{B} \boldsymbol{\Lambda} = \boldsymbol{\Psi}_{T-1}$. Thus, 
\begin{equation}
	\boldsymbol{\Psi}_T = \boldsymbol{\Psi}_{T-1} \mathbf{C}.
\end{equation}
Recalling DMD
\begin{equation}
	\boldsymbol{\Psi}_T = \mathbf{K} \boldsymbol{\Psi}_{T-1},
\end{equation}
one expects a close relationship between $\mathbf{C}$ and $\mathbf{K}$. Indeed \citep[Lemma 5.2]{krake2019dynamic}
\begin{equation}\label{eqn:CandA}
	\mathbf{C} = \big(\mathbf{V}\boldsymbol{\Sigma}^{-1} \big) \mathbf{K} \big( \boldsymbol{\Sigma} \mathbf{V}^*\big),
\end{equation}
where $\mathbf{V}$ and $\boldsymbol{\Sigma}$ are obtained from the reduced SVD of $\boldsymbol{\Psi}_{T-1}$. 

The eigenvalues of $\mathbf{K}$ are used to build the dynamics matrix $\boldsymbol{\Lambda}$. By similarity, these are the same as the eigenvalues of $\mathbf{C}$. Similar to $\mathbf{K}$, $\mathbf{C}$ is the best-fit shift operator:
\begin{equation}\label{eqn:C}
	\mathbf{C} = \underset{\mathbf{A}}{\arg\min}~ \| \boldsymbol{\Psi}_T - \boldsymbol{\Psi}_{T-1} \mathbf{A} \|_{\rm F},
\end{equation}
but we only seek the last column of $\mathbf{C}$, which we denote as $\mathbf{c} = [c_0 ~ \cdots ~ c_{T-1}]^{\top}$. Our problem simplifies to
\begin{equation}\label{eqn:c}
	\mathbf{c} = \underset{\mathbf{z}}{\arg\min}~ \| \Psi(\mathbf{x}_T) - \boldsymbol{\Psi}_{T-1} \mathbf{z} \|_2 = \boldsymbol{\Psi}_{T-1}^+ \Psi(\mathbf{x}_T).
\end{equation}

To find the modes $\mathbf{B}$ and dynamics $\boldsymbol{\Lambda}$ matrices, we first determine $\widetilde{\mathbf{K}}$ from a reduced version $\mathbf{C}$, denoted $\widetilde{\mathbf{C}}$, which can be found using a reduced SVD of $\mathbf{C}$. Then
\begin{equation}\label{eqn:AandC}
	\widetilde{\mathbf{K}} = \big( \boldsymbol{\Sigma} \mathbf{V}^*\big) \widetilde{\mathbf{C}}  \big(\mathbf{V}\boldsymbol{\Sigma}^{-1} \big)
\end{equation}
and we proceed to apply the procedure outlined in Sec.~\ref{ssec:dmd} culminating in the eigenvalues $\boldsymbol{\lambda}$ and the (eigen)modes $\mathbf{B}$. The dynamics matrix is simply $\boldsymbol{\Lambda} = [\boldsymbol{\lambda}^0 ~ \cdots ~ \boldsymbol{\lambda}^{T-1}]$, 
where the powers are taken element-wise.

We determine the amplitudes $\boldsymbol{\phi} = [\phi_0 ~ \cdots ~ \phi_{T-1}]$ by solving a least-squares problem
\begin{equation}\label{eqn:amplitudes}
	\boldsymbol{\phi} = \underset{\mathbf{z}}{\arg\min}~  \| \Psi(\mathbf{x}_1) - \mathbf{B} \mathsf{diag}(\boldsymbol{\lambda})\mathbf{z} \|_2 = ( \mathbf{B} \mathsf{diag}(\boldsymbol{\lambda}))^+ \Psi(\mathbf{x}_1)
\end{equation}
and finally, $\boldsymbol{\Phi} = \mathsf{diag}(\boldsymbol{\phi})$. Algorithm~\ref{alg:DMD} provides a summary of our implementation of DMD.

\algrenewcommand\algorithmicrequire{\textbf{Input:}}
\algrenewcommand\algorithmicensure{\textbf{Output:}}
\algrenewcommand\algorithmicprocedure{\textbf{Procedure}}
\algrenewcommand\algorithmicreturn{\textbf{Return:}}
\algrenewcommand\algorithmicend{\textbf{End}}
\algrenewcommand\algorithmicif{\textbf{If}}
\algrenewcommand\algorithmicwhile{\textbf{While}}
\algrenewcommand\algorithmicfor{\textbf{For}}
\begin{algorithm}
	\small
	\caption{Dynamic Mode Decomposition (DMD)}
	\label{alg:DMD}
	\begin{algorithmic}[1]
		\Require Data snapshots $\boldsymbol{\Psi}_{T-1} = [\Psi(\mathbf{x}_0) \cdots \Psi(\mathbf{x}_{T-1})]$ and $\boldsymbol{\Psi}_T = [\Psi(\mathbf{x}_1) \cdots \Psi(\mathbf{x}_T)]$
		\Ensure DMD modes $\mathbf{B}$, amplitudes $\boldsymbol{\Phi}$, and dynamics $\boldsymbol{\Lambda}$
		\Procedure{$\mathtt{DMD}$}{$\boldsymbol{\Psi}_{T-1},\boldsymbol{\Psi}_T$}
		\State $[\mathbf{U}, \boldsymbol{\Sigma}, \mathbf{V}] \mapsfrom \mathtt{SVD}(\boldsymbol{\Psi}_{T-1})$ \Comment{\texttt{SVD of} $\boldsymbol{\Psi}_{T-1}$}
		\State $\mathbf{c} \mapsfrom \boldsymbol{\Psi}_{T-1}^+ \Psi(\mathbf{x}_T)$ \Comment{\texttt{Companion vector}}
		\State  $\mathbf{C} \mapsfrom \left[ \begin{matrix} 0 & \cdots & 0 \\ 1 & \cdots & 0 \\ \vdots & \ddots & \vdots \\ 0 & \cdots & 1 \end{matrix} \quad \begin{matrix} \vline \\ \vline \\ \mathbf{c} \\ \vline \\ \vline \end{matrix} ~\right]$ \Comment{\texttt{Companion matrix}} 
		\State $[\mathbf{U}_{\mathbf{C}}, \boldsymbol{\Sigma}_{\mathbf{C}}, \mathbf{V}_{\mathbf{C}}] \mapsfrom \mathtt{SVD}(\mathbf{C})$ \Comment{\texttt{SVD of} $\mathbf{C}$}
		\State $\widetilde{\mathbf{C}} \mapsfrom \mathsf{P}_{\widetilde{\boldsymbol{\Sigma}}_{\mathbf{C}}}(\mathbf{C})$ \Comment{\texttt{Reduced order version of} $\mathbf{C}$} 
		\State $\widetilde{\mathbf{K}} \mapsfrom\big( \boldsymbol{\Sigma} \mathbf{V}^*\big) \widetilde{\mathbf{C}}  \big(\mathbf{V}\boldsymbol{\Sigma}^{-1} \big)$
		\State $[\widetilde{\mathbf{B}},\boldsymbol{\lambda}] \mapsfrom \mathtt{EIG}(\widetilde{\mathbf{K}})$ \Comment{\texttt{Eigenvectors/values of} $\widetilde{\mathbf{K}}$}
		\State $\mathbf{B} \mapsfrom \boldsymbol{\Psi}_T \mathbf{V} \boldsymbol{\Sigma}^{-1} \widetilde{\mathbf{B}} \mathsf{diag}(\boldsymbol{\lambda})^{-1}$ \Comment{\texttt{Modes of} $\mathbf{K}$}
		\State $\boldsymbol{\Lambda} \mapsfrom [\boldsymbol{\lambda}^0 \cdots \boldsymbol{\lambda}^{T-1}]$ \Comment{\texttt{Dynamics matrix}}
		\State $\boldsymbol{\phi} \mapsfrom (\mathbf{B} \mathsf{diag}(\boldsymbol{\lambda}))^+ \Psi(\mathbf{x}_1)$ \Comment{\texttt{Amplitudes vector}}
		\State $\boldsymbol{\Phi} \mapsfrom \mathsf{diag}(\boldsymbol{\phi})$ \Comment{\texttt{Amplitudes matrix}} \\
		\Return $\mathbf{B}$, $\boldsymbol{\Phi}$, and $\boldsymbol{\Lambda}$
		\EndProcedure
	\end{algorithmic}
\end{algorithm}

\subsection{Identification of invariant spectral features and transfer}\label{ssec:transfer}
Our method first identifies spectral features that are shared by a set of data-rich \textit{source cities} and then utilizes these features to bolster the spectral features of a target data-scarce city.

Consider a set of $m$ source cities with eigenvectors $\boldsymbol{\lambda}_1, ..., \boldsymbol{\lambda}_m$, obtained using Algorithm~\ref{alg:DMD}. 
For city $i$, the $j$th eigenvalue $\lambda_{i,j}$ describes the temporal behavior of its associated mode $\mathbf{b}_{j}$. Its magnitude $|\lambda_{i,j}|$ determines growth ($|\lambda_{i,j}| > 1$) or decay ($|\lambda_{i,j}| < 1$), and its angle (phase), when considered in polar coordinates, $\angle \lambda_{i,j}$ governs the frequency of oscillation (larger angles correspond to higher frequency oscillations). We, thus, treat similar eigenvalues for different source cities as shared spectral features.

\begin{definition}[$\epsilon$-maximally shared eigenvalues]
Let $\widehat{\boldsymbol{\lambda}} = \{\widehat{\lambda}_1, ..., \widehat{\lambda}_q\} \in \mathbb{C}^q$, where $q \le T$ and let $\epsilon > 0$ be a threshold. $\widehat{\boldsymbol{\lambda}}$ is a set of \textit{$\epsilon$-maximally shared eigenvalues} if 
\begin{enumerate}
	\item[(i)] for all $1 \le k \le q$, there exist indices $j_1,...,j_m$ such that
	\begin{equation}
		|\widehat{\lambda}_k - \lambda_{i,j_i}| < \epsilon, \mbox{ for all } i = 1,...,m, \mbox{ and}
	\end{equation}
	\item[(ii)] $q$ is the largest number of eigenvalues that satisfy condition (i).
\end{enumerate}
\end{definition}

To find a set of $\epsilon$-maximally shared eigenvalues, we simply select the eigenvector associated with any of the source cities, say city $k$, compare each of its eigenvalues, $\lambda_{kl}$ $l=1,...,T$, to all eigenvalues in the other source cities. If all other source cities have at least one eigenvalue that satisfies the difference condition, i.e., there exists an index $j_i$ so that $|\lambda_{kl} - \lambda_{i,j_i}| < \epsilon$ for all $i \ne k$, then $\lambda_{kl}$ is appended to $\widehat{\boldsymbol{\lambda}}$. Algorithm~\ref{alg:MaxShEigen} summarizes the procedure.

\begin{algorithm}
	\small
	\caption{$\epsilon$-Maximally Shared Eigenvalues}
	\label{alg:MaxShEigen}
	\begin{algorithmic}[1]
		\Require Source city eigenvectors $\{\boldsymbol{\lambda}_1, ..., \boldsymbol{\lambda}_m\}$, threshold $\epsilon$, benchmark city $k \in \{1,...,m\}$
		\Ensure $\epsilon$-Maximally shared eigenvalues $\{\widehat{\lambda}_1, ..., \widehat{\lambda}_q\}$ 
		\State $\widehat{\boldsymbol{\lambda}} \mapsfrom \emptyset$ \Comment{\texttt{Initialization}}
		\State \textbf{For} $l =1:T$
		\State \hspace{0.1 in}\textbf{If} $\max_{i \ne k} \min_j |\lambda_{k,l} - \lambda_{i,j}| < \epsilon$ \textbf{then}
		\State \hspace{0.2 in} $\widehat{\boldsymbol{\lambda}} \mapsfrom \widehat{\boldsymbol{\lambda}} \cup \lambda_{k,l}$
		\State \hspace{0.1 in}\textbf{End If}
		\State \textbf{End For}
	\end{algorithmic}
\end{algorithm}

To transfer these spectral features to a target city, we employ a constrained DMD approach \citep{krake2022constrained} to enhance the DMD of the target (data-sparse) city using the $\epsilon$-maximally shared eigenvalues determined from the source cities.  Let $\mathbf{B}$, $\boldsymbol{\Phi}$, and $\boldsymbol{\Lambda}$ denote the spectral decomposition of the target city (determined using Algorithm~\ref{alg:DMD} applied to data from the target city). We aim to calculate an enhanced companion matrix using $\widehat{\boldsymbol{\lambda}}$ for the target city in a way that preserves its spectral features. The characteristic polynomial of the companion matrix 
\begin{equation}
	p_{\mathbf{C}}(\lambda) = -c_0 - c_1 \lambda - \cdots - c_{T-1} \lambda^{T-1} + \lambda^T
\end{equation}
satisfies $p_{\mathbf{C}}(\lambda_i) = 0$ if and only if  $\lambda_i$ is an eigenvalue of $\mathbf{C}$. Hence, to preserve the original eigenvalues, $\mathbf{c}$ must satisfy
\begin{equation}
	\boldsymbol{\Lambda} \mathbf{c} = \boldsymbol{\xi},
\end{equation}
where $ \boldsymbol{\xi} = [\lambda_1^T ~ \cdots ~ \lambda_T^T]^{\top}$. 

To facilitate the transfer of $\widehat{\boldsymbol{\lambda}}$, define $\widehat{\boldsymbol{\Lambda}} \in \mathbb{C}^{q \times T}$ and $\widehat{\boldsymbol{\xi}} \in \mathbb{C}^q$ as
\begin{equation}
	\widehat{\boldsymbol{\Lambda}} \equiv \begin{bmatrix}
		1 & \widehat{\lambda}_1 & \widehat{\lambda}_1^2 & \cdots & \widehat{\lambda}_1^{T-1} \\
		\vdots & \vdots & \vdots & \ddots & \vdots \\
		1 & \widehat{\lambda}_q & \widehat{\lambda}_q^2 & \cdots & \widehat{\lambda}_q^{T-1}
	\end{bmatrix}
\mbox{ and } ~ \widehat{\boldsymbol{\xi}} \equiv \begin{bmatrix}
	\widehat{\lambda}_1^T \\ \vdots \\ \widehat{\lambda}_q^T
\end{bmatrix}.
\end{equation}
We calculate the enhanced companion vector, $\overline{\mathbf{c}}$, by solving the constrained problem
\begin{equation}\label{eqn:cEnhanced}
	\overline{\mathbf{c}} = \underset{\mathbf{z}}{\arg\min}~ \| \boldsymbol{\xi} - \boldsymbol{\Lambda} \mathbf{z} \|_2:~ \widehat{\boldsymbol{\Lambda}} \mathbf{z} = \widehat{\boldsymbol{\xi}},
\end{equation}
which has the closed form solution \cite[Appendix]{krake2022constrained}
\begin{equation}
		\overline{\mathbf{c}} = \mathbf{c} + (\boldsymbol{\Lambda}^* \boldsymbol{\Lambda})^{-1} \widehat{\boldsymbol{\Lambda}}^* \big(  \widehat{\boldsymbol{\Lambda}} (\boldsymbol{\Lambda}^* \boldsymbol{\Lambda})^{-1} \widehat{\boldsymbol{\Lambda}}^*  \big)^{-1}(\widehat{\boldsymbol{\xi}} - \widehat{\boldsymbol{\Lambda}} \mathbf{c}).
\end{equation}
To obtain the enhanced decomposition, we simply replace $\mathbf{c}$ with $\overline{\mathbf{c}}$ in Algorithm~\ref{alg:DMD}. 

\subsection{Henkelized dynamic mode decomposition (HDMD)}
For city traffic data, the number of rows in $\boldsymbol{\Psi}_T$ corresponds to the number of sensors in the system (be they fixed or mobile). This tends to be a small number when compared with the dimensionality of fluid systems for which DMD techniques continue to show success. 

In order to discover latent features, and specifically for the purpose of temporal prediction, we need a higher-dimensional system (rows). To this end, we employ a powerful technique known as \textit{time-delay embedding} \citep{kutz2016dynamic}, which involves the Hankel transformation ($\mathcal{H}$) of our original data matrices using a \textit{delay parameter} $h$:
\begin{multline}
		\boldsymbol{\Psi}_{T-1} = \begin{bmatrix}
		\Psi_0 & \cdots & \Psi_{T-1}
	\end{bmatrix}
	\\ \mapsto 
	\mathcal{H}_h \boldsymbol{\Psi}_{T-1} = \begin{bmatrix}
		\Psi_0 & \cdots & \Psi_{T-h} \\
		\Psi_1 & \cdots & \Psi_{T-h+1} \\
		\vdots & \ddots & \vdots \\
		\Psi_h & \cdots & \Psi_{T-1}
	\end{bmatrix} 
\end{multline}
and $\mathcal{H}_h \boldsymbol{\Psi}_{T-1} \in \mathbb{R}^{ht \times (T-h)}$.

\begin{mdframed}[style=myStyle]
	
	\smallskip
	
    The key intuition behind the embedding is 
    we assume that the dynamic behavior of the observables depends on latent variables. This dependency, in turn, \textit{encodes features of the hidden variables in the observables}. Indeed, if a variable $\alpha$ depends on another (hidden) variable $\beta$, then there exists a function $f$ such that $\alpha = f(\beta)$. 
    
    Theoretically, the method is rooted in Takens' embedding theorem \citep{takens2006detecting}, which suggests that the dynamics of a system can be comprehensively captured by embedding its output into a higher dimensional space using time delays. Essentially, using delayed values of the observed data as proxies for unmeasured state variables, time-delay embedding unfolds the underlying dynamics that might be obscured in lower-dimensional observations. Our Hankel matrix embedding is akin to \emph{watching scenes in a movie many times (with rewinding)} with the aim of uncovering hidden details.

    In our context, where city data exhibit periodic behavior, for sufficiently large $h$, the top $j$ right-singular vectors\footnote{Recall that the $\mathbf{V}$ encodes the dynamical features of the data matrix.} $\mathbf{v}_1, ..., \mathbf{v}_j$ (obtained from the SVD $\mathbf{U} \boldsymbol{\Sigma} \mathbf{V}^* = \mathcal{H}_h \boldsymbol{\Psi}_{T-1}$) should be the simplest periodic functions, namely sine and cosine waves. This is indicative of an embedding that \emph{discovers} the latent features, thereby linearizing the system.
    
    \smallskip
    
\end{mdframed}

To determine the spectral features of the source cities, we apply the DMD algorithm (Algorithm~\ref{alg:DMD}) to Hankelized inputs. Similarly, the enhanced DMD uses Hankelized inputs for the source and target cities. To emphasize the time-delay embedding, we will henceforth refer to our methods as \textit{Henkelized Dynamic Mode Decomposition} (HDMD) and \textit{Enhanced HDMD via Transfer} (TrHDMD). 
We summarize the TrHDMD procedure in Algorithm~\ref{alg:CDMD}. 
\begin{algorithm}
	\small
	\caption{Enhanced HDMD via Transfer (TrHDMD)}
	\label{alg:CDMD}
	\begin{algorithmic}[1]
		\Require Target City Data snapshots $\boldsymbol{\Psi}_{T-1} = [\Psi(\mathbf{x}_0) \cdots \Psi(\mathbf{x}_{T-1})]$ and $\boldsymbol{\Psi}_T = [\Psi(\mathbf{x}_1) \cdots \Psi(\mathbf{x}_T)]$, $\epsilon$-maximally shared eigenvectors $\widehat{\boldsymbol{\lambda}}$, Delay $h$
		\Ensure Enhanced DMD modes $\overline{\mathbf{B}}$, amplitudes $\overline{\boldsymbol{\Phi}}$, and dynamics $\overline{\boldsymbol{\Lambda}}$
		\Procedure{$\mathtt{TrDMD}$}{$\boldsymbol{\Psi}_{T-1},\boldsymbol{\Psi}_T,\widehat{\boldsymbol{\lambda}}$}
            \State $[\mathbf{B}, \boldsymbol{\Phi}, \boldsymbol{\Lambda}] \mapsfrom \mathtt{DMD}(\mathcal{H}_h \boldsymbol{\Psi}_{T-1},\mathcal{H}_h \boldsymbol{\Psi}_T)$ \Comment{\texttt{Record} $[\mathbf{U}, \boldsymbol{\Sigma}, \mathbf{V}]$ \texttt{from} $\mathtt{SVD}(\mathcal{H}_h \boldsymbol{\Psi}_{T-1})$ \texttt{Step}}
            \State $\mathbf{c} \mapsfrom \boldsymbol{\Psi}_{T-1}^+ \Psi(\mathbf{x}_T)$
            \State $\widehat{\boldsymbol{\Lambda}} \mapsfrom \begin{bmatrix}
		1 & \widehat{\lambda}_1 & \widehat{\lambda}_1^2 & \cdots & \widehat{\lambda}_1^{T-1} \\
		\vdots & \vdots & \vdots & \ddots & \vdots \\
		1 & \widehat{\lambda}_q & \widehat{\lambda}_q^2 & \cdots & \widehat{\lambda}_q^{T-1}
	\end{bmatrix}
\mbox{ and } ~ \widehat{\boldsymbol{\xi}} \mapsfrom \begin{bmatrix}
	\widehat{\lambda}_1^T \\ \vdots \\ \widehat{\lambda}_q^T
\end{bmatrix}$
            \State $\overline{\mathbf{c}} \mapsfrom \mathbf{c} +(\boldsymbol{\Lambda}^* \boldsymbol{\Lambda})^{-1} \widehat{\boldsymbol{\Lambda}}^* \big(  \widehat{\boldsymbol{\Lambda}} (\boldsymbol{\Lambda}^* \boldsymbol{\Lambda})^{-1} \widehat{\boldsymbol{\Lambda}}^*  \big)^{-1}(\widehat{\boldsymbol{\xi}} - \widehat{\boldsymbol{\Lambda}} \mathbf{c})$
            \State  $\mathbf{C} \mapsfrom \left[ \begin{matrix} 0 & \cdots & 0 \\ 1 & \cdots & 0 \\ \vdots & \ddots & \vdots \\ 0 & \cdots & 1 \end{matrix} \quad \begin{matrix} \vline \\ \vline \\ \overline{\mathbf{c}} \\ \vline \\ \vline \end{matrix} ~\right]$ \Comment{\texttt{Companion matrix}}
            \State $[\mathbf{U}_{\mathbf{C}}, \boldsymbol{\Sigma}_{\mathbf{C}}, \mathbf{V}_{\mathbf{C}}] \mapsfrom \mathtt{SVD}(\mathbf{C})$ 
		\State $\widetilde{\mathbf{C}} \mapsfrom \mathsf{P}_{\widetilde{\boldsymbol{\Sigma}}_{\mathbf{C}}}(\mathbf{C})$ 
		\State $\widetilde{\mathbf{K}} \mapsfrom\big( \boldsymbol{\Sigma} \mathbf{V}^*\big) \widetilde{\mathbf{C}}  \big(\mathbf{V}\boldsymbol{\Sigma}^{-1} \big)$
		\State $[\widetilde{\mathbf{B}},\boldsymbol{\lambda}] \mapsfrom \mathtt{EIG}(\widetilde{\mathbf{K}})$ 
		\State $\overline{\mathbf{B}} \mapsfrom \boldsymbol{\Psi}_T \mathbf{V} \boldsymbol{\Sigma}^{-1} \widetilde{\mathbf{B}} \mathsf{diag}(\boldsymbol{\lambda})^{-1}$ 
		\State $\overline{\boldsymbol{\Lambda}} \mapsfrom [\boldsymbol{\lambda}^0 \cdots \boldsymbol{\lambda}^{T-1}]$ 
		\State $\overline{\boldsymbol{\Phi}} \mapsfrom (\overline{\mathbf{B}} \mathsf{diag}(\boldsymbol{\lambda}))^+ \Psi(\mathbf{x}_1)$ 
		\State $\boldsymbol{\Phi} \mapsfrom \mathsf{diag}(\boldsymbol{\phi})$ 
		\EndProcedure
	\end{algorithmic}
\end{algorithm}

\section{Experiments}\label{sec:experiments}
\subsection{The data}
We utilize the B22 dataset, a refined iteration of the UTD19 dataset \citep{loder2019understanding,bramich2022fitting,bramich2023fitfun}, which aggregates data from 10,150 loop detectors across 25 cities, resulting in 168,183,675 flow-occupancy pairs, including 147,034,646 reliable pairs without error flags. The dataset also includes subsets of flow-speed and flow-occupancy-speed measurements. Each data pair or triple reflects an aggregation over a 3 to 5-minute interval. Given the dataset’s multi-source origins, typically local city authorities, there is notable heterogeneity in data quantity, quality, and temporal coverage across different cities. The dataset is available for download at the Harvard Dataverse \citep{DVN/FSGDQM_2022}.

Our experiments examine four cities —Graz, Bordeaux, Madrid, and Stuttgart— selected from the B22 dataset. The cities' datasets feature extended records over consecutive days ($\ge 5$ days) with a consistent sampling interval of 5 minutes, making them ideal for our experiments. For each of the selected cities, we randomly sampled 30 detectors and four consecutive days to construct a multivariate time series. We selected 30 detectors per city to accommodate the variable detector counts in the B22 dataset. Additionally, for data consistency, we chose mid-week workdays for the source cities, while the target city's data included both workdays and off-days.

In the experiment, the first three days' data were allocated for training purposes, while the fourth day's data were kept for testing prediction accuracy. Conceptually, this multivariate training time series can be viewed as a spatiotemporal matrix of size $30\times 864$, where rows represent detectors and columns represent time stamps, with each cell indicating a flow rate over the corresponding spatiotemporal cell. As a preprocessing step, we centered the data by subtracting the time-averaged flows from the data to enhance analysis accuracy 
\citep{avila2020data,mezic2005spectral,mezic2020spectrum}.  

In our experimental setup, we randomly selected loop detectors deployed across cities to capture distinct traffic patterns. Spectral characteristics were identified using HDMD based on aggregated flow readings. 
Koopman methods are also applicable to disaggregated data: Avila and Mezi{\'c} \citep{avila2020data} have demonstrated the Koopman method's effectiveness in uncovering intricate spatiotemporal patterns using NGSIM US101 data. For instance, they successfully validated phenomena like the ``pumping effect,'' where traffic wave amplitudes decrease upstream past on-ramps and increase past off-ramps. The modes also reveal other patterns, such as moving localized clusters, such as those reported in \cite{kaufmann2018aerial} and stop-and-go waves. 

\subsection{Embedding and linearization tests}\label{sec:exp-linearization}
\begin{figure}[!]
\centering
\includegraphics[width=.44\textwidth]{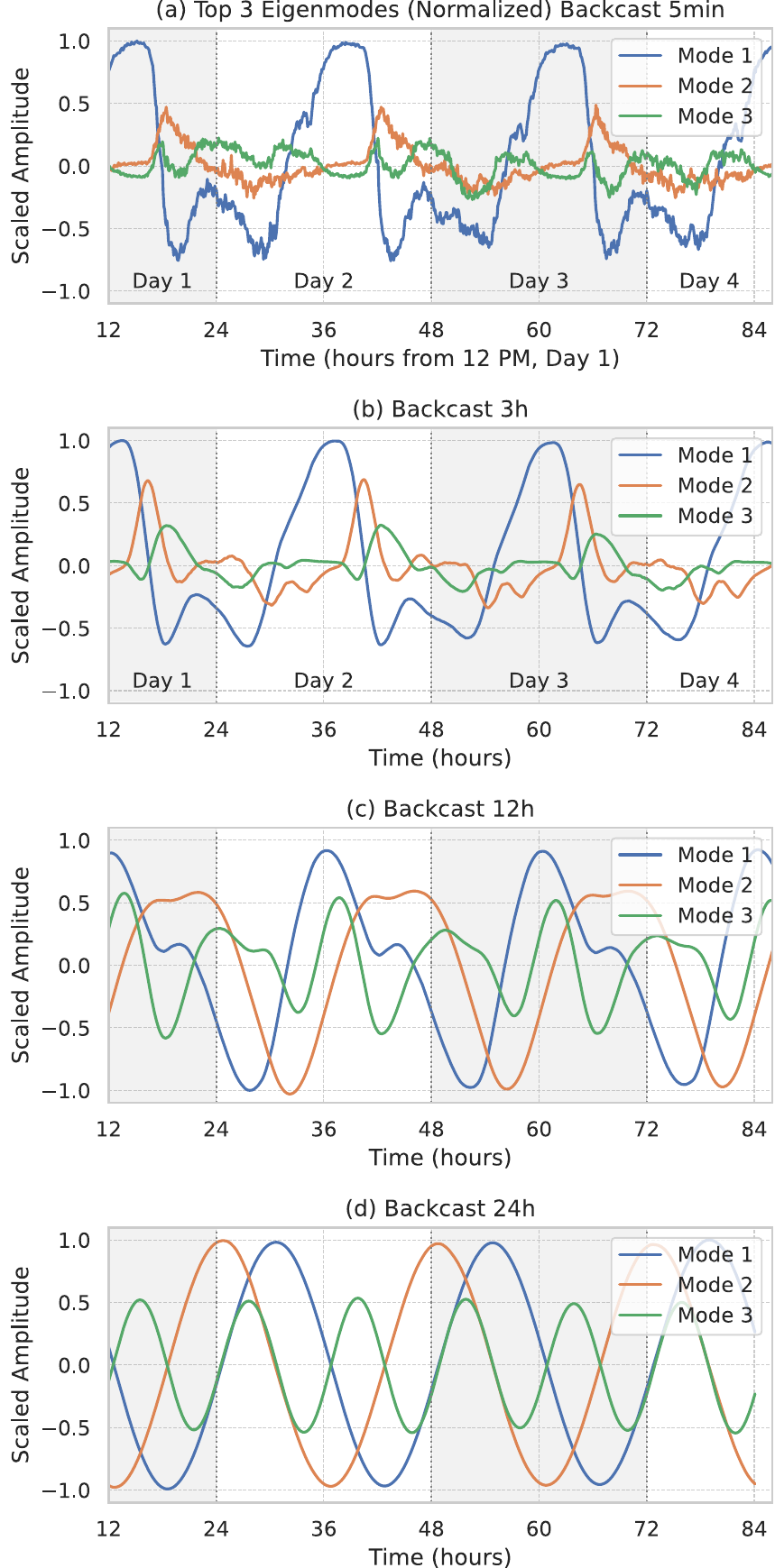}
	\caption{Visualization of Averaged Temporal Dynamics in Stuttgart Traffic via Koopman Operator Analysis: this figure presents the top three eigenmodes extracted from right-singular eigenvectors of Stuttgart traffic data, spanning several days and depicted in both the original observational space and the transformed high-dimensional Koopman space. Each subplot (a through d) displays the normalized singular vectors corresponding to varied backcasting intervals: 5 minutes, 3 hours, 12 hours, and 24 hours, respectively. These eigenmodes are scaled with their eigenvalues and are maximally normalized within the range of [-1, 1]. They exhibit periodic behaviors that underscore the traffic patterns across the Stuttgart region. The chosen delay or backtest period influences the complexity of the observed patterns: initially, the traffic dynamics demonstrate noisy, complex, high-frequency, non-linear characteristics. Yet, within the expansive Koopman space, these dynamics are simplified into basic quasi-sinusoidal waveforms. The selection of the delay/backcast window balances between promoting linearity and preserving the informative nature of the eigenmodes.}
	\label{fig:DMDSvd}
\end{figure}

Our first experiment examines the Koopman method's ability to generate a Fourier basis for urban traffic dynamics using time-delay embedding. In other words, we aim to see if an appropriate choice of the parameter $h$ results in a simple system of right-singular eigenvectors (namely, sine and cosine waves).  In Fig.~\ref{fig:DMDSvd}, we display the top four right-singular vectors of Stuttgart’s time snapshot matrix, $\boldsymbol{\Psi}_{T-1}$ without embedding (Fig.~\ref{fig:DMDSvd}a), with $h=250$ (Fig.~\ref{fig:DMDSvd}b), and with $h = 300$ (Fig.~\ref{fig:DMDSvd}c). We focus on right-singular vectors $\mathbf{V}$, as they contain temporal information essential for reconstructing $\boldsymbol{\Psi}_{T-1}$. By analyzing $\mathbf{V}$, we uncover dynamic patterns within the traffic system through Singular Value Decomposition (SVD), which provides a hierarchical time series that defines the traffic dynamics \citep{brunton2019notes}.

The illustrations reveal distinct behaviors: the initial graph of $\boldsymbol{\Psi}_{T-1}$ shows highly nonlinear, noisy vectors, reflecting complex traffic patterns. The subsequent graphs with embedding delays $h=250$ and $h=300$ progressively exhibit more regular, sinusoidal patterns. This transition illustrates that increasing the delay parameter \textit{linearizes} the traffic dynamics by projecting them into a higher-dimensional space. 
In summary, for Stuttgart, with $h=300$, we \textit{discovered} the dimension of the latent space that linearizes the system. In this case, the nonlinear ``spatial'' dynamics of the $t=30$ sensors live in a linear space that is $ht = 9000$ dimensional.

\subsection{Eigenvalues and shared modes}\label{sec:exp-eigen-extract}

\begin{figure}[h!]
    \centering
\includegraphics[width=.50\textwidth]{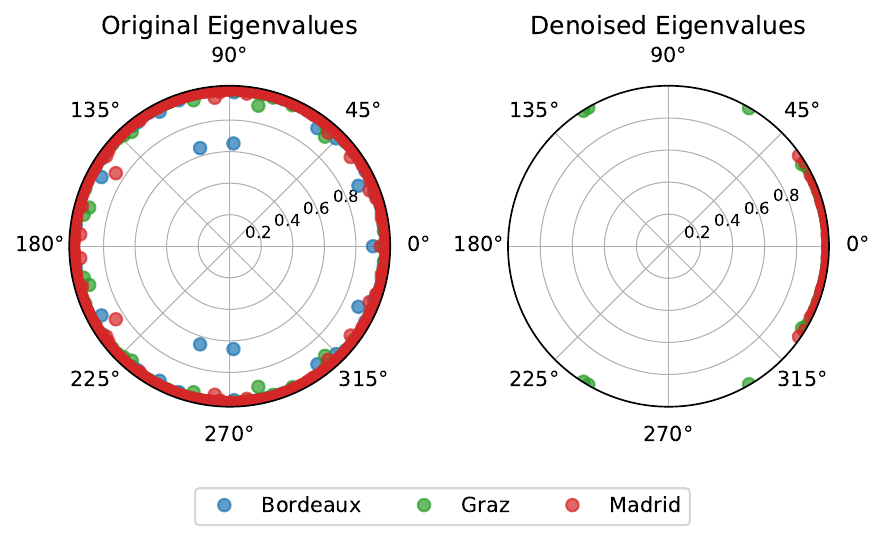}%
\caption{Polar comparison of Koopman eigenvalues among source cities: (a) Displays all eigenvalues calculated by HDMD with a delay of 300 for Bordeaux, Graz, and Madrid, based on data trained over three consecutive days. (b) For enhanced clarity, only eigenvalues exceeding an amplitude threshold of $10^{-2}$ are shown, determined using an optimal singular value hard thresholding (SVHT) procedure \citep{dl2014optimal}. There is significant overlap among the eigenvalues on the unit circle, indicating consistent, shared patterns across the cities. Notably, the symmetry along the $ \angle \lambda=0 $ axis illustrates that the eigenvalues are pairwise conjugated, as they originate from the diagonalization of a real matrix.}

	\label{fig:eigenvalueDisplay}
\end{figure}

The eigenvalues, depicted in Fig.~\ref{fig:eigenvalueDisplay}, provide insights into fundamental modes or patterns preserved as the system evolves in time, with their magnitudes indicating the growth or decay rate of a mode ($|\lambda|$) and their angles ($\angle \lambda$) indicating the associated mode's oscillation frequency, which we will represent using the cycle time of the eigenvalue below. 

 The top polar chart in Fig.~\ref{fig:eigenvalueDisplay} provides an illustration of \textit{all} eigenvalues from the three source cities. The majority lie exactly on the unit circle, signifying stable patterns that are time-independent (no growth or decay in the associated mode). We do not observe eigenvalues outside the unit circles, indicating that we have discovered stable time-independent patterns in the data; the minority that lies in the interior of the unit circle corresponds to decaying modes, which means that they are ultimately short-lived. 

 Fig.~\ref{fig:eigenvalueDisplay}b presents the eigenvalues that remain after applying an optimal singular value hard thresholding (SVHT) procedure \citep{dl2014optimal}. The clusters in the unit circle's first and fourth quadrants indicate patterns consistent across the three cities.

To identify shared patterns across cities in the form of a set of $\epsilon$-Maximally Shared Eigenvalues, 
we set $\epsilon$ to 0.001. We use the \textit{cycle time} associated with each eigenvalue as a measure of periodicity. The cycle time, $\sigma$, of eigenvalue $\lambda$ is calculated as
\begin{equation}
	\sigma = \frac{2\pi \Delta t}{ \mathrm{Im}(\log(\lambda)) },
\end{equation}
where $\Delta t$ is the sampling time interval. 
With $h=300$, we identify a total of 7 shared cycle times: [0.53, 2.38, 4.81, 5.98, 7.99, 12.02, 24.21] (unit: hr), and with $h = 250$, a total of 10 shared cycle times emerge [1.14, 1.20, 2.39, 3.00, 3.44, 4.81, 6.00, 7.99, 12.02, 24.17] (unit: hr). This set of shared cycle times encompasses not only common daily patterns (24hr) and day-night patterns (12hr), indicating promising progress in pattern identification but also capturing smaller periodic fluctuations. Comparison between the two sets reveals that nearly all cycle times from $h=300$ are included in the set of $h = 250$, with discrepancies primarily concerning smaller periodic fluctuations (cycle times $\leq 4$ hrs). These fluctuations may be attributed to various factors, such as scheduled arrivals and departures of public transportation systems or commercial activities. 

In summary, we have decomposed the traffic dynamics of the source cities into $ht$ sub-patterns $\mathbf{b}_j$, $j=1,...,ht$ for each city $m$. Each sub-pattern $j$ is associated with a frequency, i.e. $\lambda_j^k \phi_j(\mathbf{x}_0) \mathbf{b}_j$ for $k = 0,1,2,...$ from \eqref{eqn:KoopmanOnFObs}. We have identified the sub-patterns that share the same frequency across all source cities, where $\widehat{\boldsymbol{\lambda}} \subseteq \boldsymbol{\lambda}_m $.

\begin{figure}[!]
	\centering
\includegraphics[width=.5\textwidth]{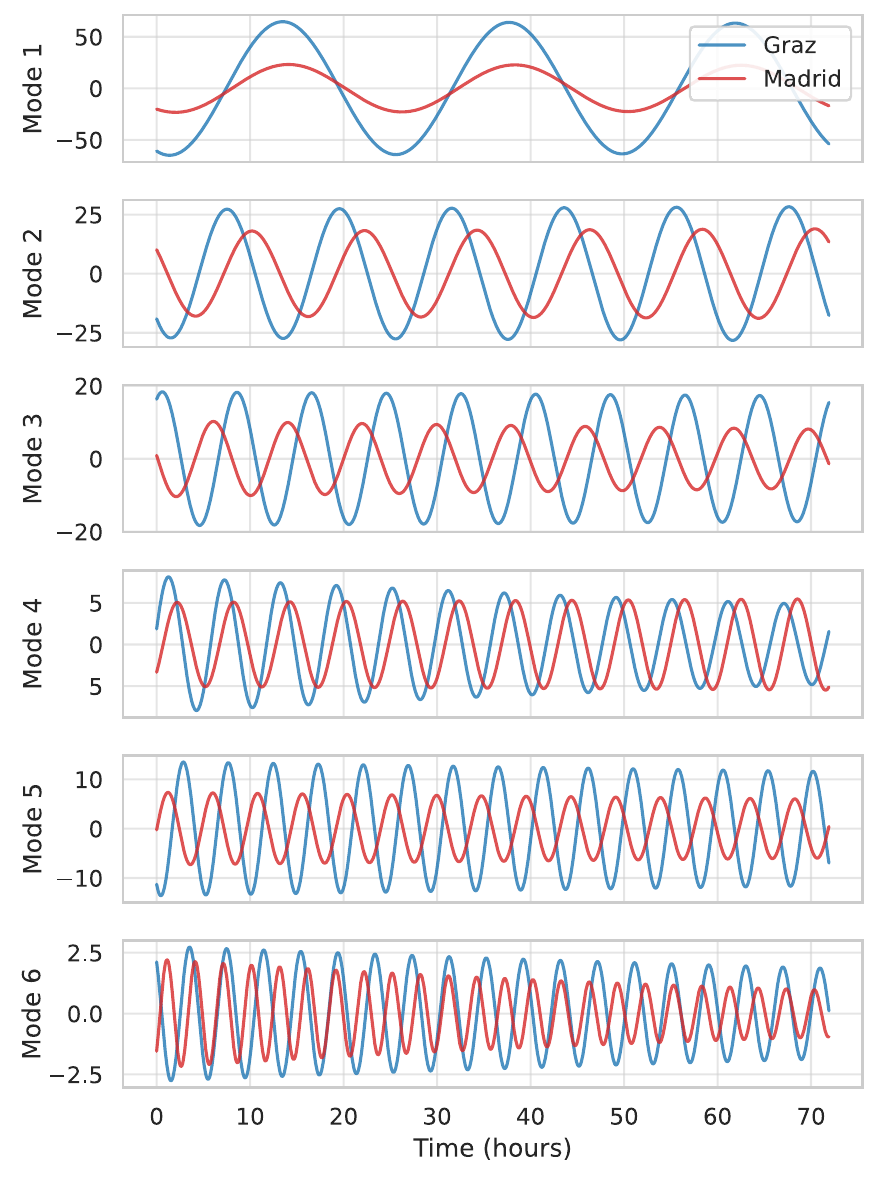}%

	\caption{Subpatterns with shared frequencies in Graz and Madrid. We observe that they still differ in both amplitude and phase. Within the Koopman framework, the cross-city shared, and transferrable information is identified from the shared Koopman eigenvalues, while each city's unique characteristics are preserved in both the city-specific Koopman modes and the Koopman eigenvalues.}
	\label{fig:shared-subpatterns-comparison}
\end{figure}

It is important to note that there is still a large number of city-specific, i.e., not shared, sub-patterns. Readers familiar with the cities used in this study may recognize that their daily schedules differ; for instance, Madrid's schedule tends to peak much later both during the morning and evening peaks than Graz's, as Spanish cities often have a slightly later start to the day compared to many Central European cities.  As illustrated in Figure~\ref{fig:shared-subpatterns-comparison}, where we randomly select one loop detector from Graz and one from Madrid, we plot the 7 sub-patterns associated with the identified cycle times of ($\lambda_j \in \widehat{\boldsymbol{\lambda}}$).

\begin{figure*}[!]
\centering
\includegraphics[width=1\textwidth]{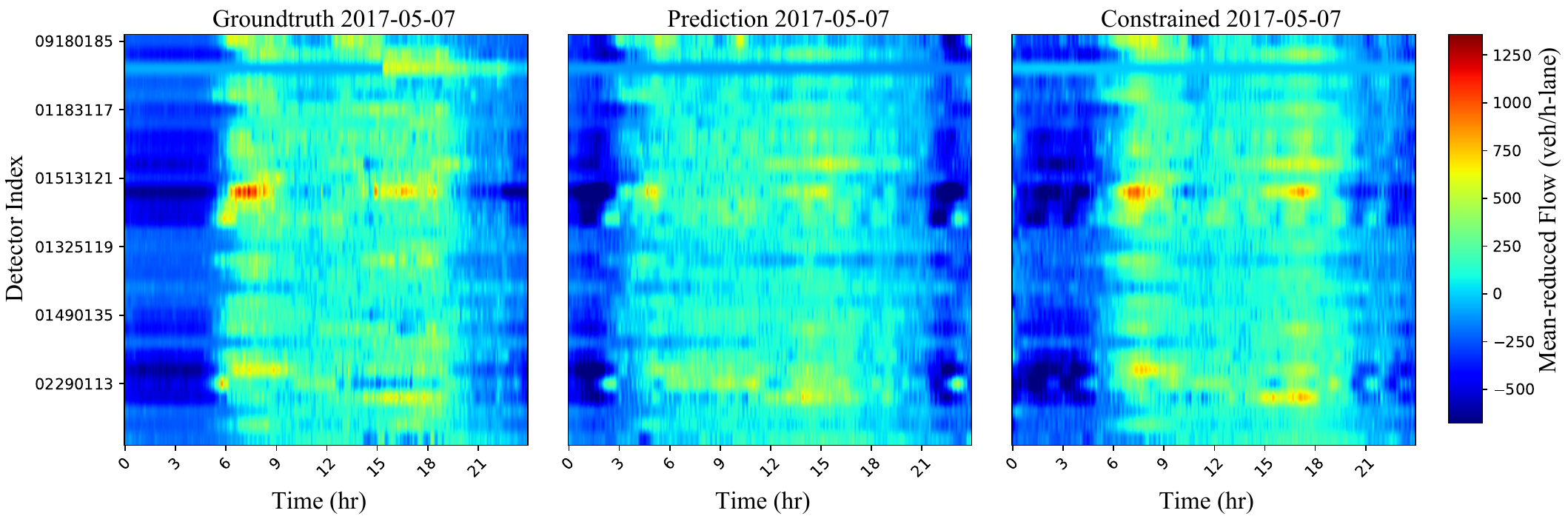}

\caption{Comparison of HDMD and TrHDMD prediction results. The graphs show flow readings of 30 detectors in the target city (Stuttgart) on May 07, 2012. The left graph is the ground truth, the middle graph shows predicted flows using HDMD, and the right graph illustrates the TrHDMD predictions. We observe that TrHDMD has a better prediction performance, with the congested periods and free-flow periods more accurately captured than with HDMD. }
\label{fig:5-min-test}
\end{figure*}

The illustrated sub-patterns are sorted in decreasing order of period from front to back. The time intervals of these sub-patterns are identical to the training interval, starting from midnight on a workday and ending 3 days later. It is evident that patterns with the same cycle time still vary in both amplitude and phase. For instance, the 24-hour cycle time pattern in Graz has a greater amplitude than in Madrid. We also notice that the smaller periodic fluctuations are not stable and decay in both cities. In summary, within the Koopman framework for city traffic systems, we identify the shared fundamental frequencies from the shared Koopman eigenvalues, while each city's unique characteristics are preserved in both the Koopman modes and the Koopman eigenvalues unique to them.

\subsection{Overall prediction performance of HDMD and TrHDMD}\label{sec:pred-comparison}
Similar to our previous experiments, we selected 30 detectors from Stuttgart randomly to assess the predictive performance of HDMD and TrHDMD. Again, we train our model using the first three days' data and predict the fourth day. The results show that TrHDMD outperforms HDMD, which suggests the effective transferability of shared frequencies from source cities. Fig.~\ref{fig:5-min-test} illustrates this on May 07, 2017, using data from the previous day.\footnote{We only display 6 detector names from all 30 detectors on the vertical axis for aesthetic reasons. The 30 detectors are ordered randomly on the vertical axis. The order does not reflect the relative locations of the detectors in the network.} The heatmap displays mean-reduced flow readings, highlighting TrHDMD's improved accuracy in identifying congestion (red) and free-flow periods (dark blue) compared to HDMD.

For a more comprehensive assessment, we compare HDMD and TrHDMD with state-of-the-art methods used in this context, namely multivariate time series methods, which range from naive baseline models like naive/autoregressive moving average and classical approaches such as vector autoregressive moving average (VARIMA), the Kalman filter, RandomForest, XGBoost, to neural network based models like neural basis expansion analysis time series forecasting (N-BEATS) \citep{oreshkin2019n}, Long short-term memory (LSTM), and time series data generative models like Timeseries DGAN \citep{lin2020using}. Table \ref{tab:alg-comparison} presents a summary of the results obtained using 30 detector readings over three days from Stuttgart for training and predicting the following day.
\begin{table}[!ht]
\caption{Multivariate time series methods performance comparison. All algorithms use 3 days of input from 30 randomly selected detectors in Stuttgart for training and predict the readings of the following day. TrHDMD exhibits competitive performance, closely approaching the optimal results across all metrics. Moreover, TrHDMD demonstrates improvement over HDMD, indicating the positive impact of transferring shared frequencies from source cities. \label{tab:alg-comparison}}
\centering
\begin{tabular}{|l|l|l|l|l|}
\hline
Method & RE & MAE & DTW & CS\\
\hline
\hline
MA & 0.9816  & 206.32567  & 23014 & 0.1745\\

\hline
VARIMA & 0.9021  & 186.2237  & 21484 & 0.4354\\
\hline
Kalman filter & 0.7066 & 137.9796  & 16999 & 0.7148\\

\hline
RandomForest & 0.5570 & 110.3526 & \textbf{9385} & 0.8055 \\
\hline
XGBoost & 0.4746 & 83.0778 & 10259 & 0.8844\\

\hline
N-BEATS & \textbf{0.4118} & \textbf{74.9470} & 10023 & \textbf{0.9081}\\
\hline
LSTM & 1.0049 & 220.6972 & 25045 & -0.0601\\
\hline
Timeseries DGAN & 1.4768 & 351.2057 &  32766 & 0.0662\\

\hline
HDMD (h=300) &  0.4411 & 78.3157 & 10855 & 0.8861\\

TrHDMD (h=300) & 0.4399 & 78.0011 & 10861 & 0.8861\\

\hline
HDMD (h=250) &  0.4454 & 80.4433 & 10925 & 0.8852\\

TrHDMD (h=250) & 0.4307 & 78.0500 & 10617 & 0.8949\\

\hline
\end{tabular}
\end{table}

When evaluating the quality of multivariate time series prediction, we consider multiple factors and concepts such as alignment of trend, seasonality, and frequency/angular preservation \citep{fernandez2023similarityts}. Therefore, in addition to standard relative error (RE) calculations, we incorporate additional similarity measures commonly used in multivariate time series similarity analysis. These measures include the multidimensional dynamic time warping (DTW) \citep{shokoohi2017generalizing,meert_2022_7158824} and row-average cosine similarity (CS) as well as more traditional metrics like mean absolute error (MAE). We describe these metrics below: 
\begin{itemize}
\item For two matrices $\widehat{\mathbf{Y}}$ and $\mathbf{Y}$, RE is calculated as 
\begin{equation}
    \text{RE} = \frac{\| \widehat{\mathbf{Y}} - \mathbf{Y} \|_{\rm F}}{\|\mathbf{Y}\|_{\rm F}}.
\end{equation}
\item CS is a measure of similarity between two non-zero vectors of an inner product space, computed as the dot product of the vectors divided by the product of their lengths. It measures similarity in shape independent of amplitude or offset. One can also interpret CS as a measure of the correlation between two matrices $\widehat{\mathbf{Y}}$ and $\mathbf{Y}$. It is calculated as 
    \begin{equation}
        \mathrm{CS} = \frac{1}{|\mathbf{Y}_{:,1}|} \sum_{i=1}^{|\mathbf{Y}_{:,1}|}  \frac{\langle\widehat{\mathbf{Y}}_{i,:} , \mathbf{Y}_{i,:} \rangle}{\|\widehat{\mathbf{Y}}_{i,:}\|_2 \|\mathbf{Y}_{i,:}\|_2},
    \end{equation}
    where $|\mathbf{Y}_{:,1}|$ is the size of the vector $\mathbf{Y}_{:,1}$ (number of columns of $\mathbf{Y}$) and $\|\cdot\|_2$ is the $\ell_2$ vector norm. We average over rows of $\widehat{\mathbf{Y}}$ and $\mathbf{Y}$ because each row represents an individual detector's reading over the predicted time interval. 
\item DTW assesses how well the temporal relationship was maintained between the two time series by aligning them in a way that minimizes the distance between them. DTW is implemented by applying dynamic programming to fill a matrix $\mathbf{W}$, whose entry $\mathbf{W}_{i,j}$ represents the minimum cumulative distance required to align the sub-series $\mathbf{Y}_{:,:i}$ and $\widehat{\mathbf{Y}}_{:,:j}$\footnote{$\mathbf{Y}_{:,:i}$ is a matrix that consists of the first $i$ columns of $\mathbf{Y}$.}, calculated as
    \begin{equation}
        \mathbf{W}_{i,j} = \mathbf{D}_{i,j} + \min \{ \mathbf{W}_{i-1,j},\mathbf{W}_{i,j-1},\mathbf{W}_{i-1,j-1}\},
    \end{equation}
    where $\mathbf{D}_{i,j}$ is the Euclidean distance between $\mathbf{Y}_{:,i}$ and $\widehat{\mathbf{Y}}_{:,j}$. We report $\mathbf{W}_{n,n}$ to demonstrate the optimal alignment of the predicted time series and the ground truth.
\item MAE measures the average magnitude of errors between the predicted and actual values. It is calculated as follows:
    \begin{equation}
        \mathrm{MAE} = \frac{1}{|\mathbf{Y}_{:,1}| \times |\mathbf{Y}_{1,:}|}\sum_{i=1}^{|\mathbf{Y}_{:,1}|} \sum_{j=1}^{|\mathbf{Y}_{1,:}|} |\mathbf{Y}_{i,j} - \widehat{\mathbf{Y}}_{i,j}|.
    \end{equation}
\end{itemize}

In summary, a smaller RE/MAE value represents a more accurate prediction, measured in a relative and absolute magnitude, respectively. Likewise, a lower DTW value suggests better temporal alignment between the predicted time series and the ground truth. Finally, CS falls within the interval $[-1,1]$, where 1 means the compared time series are perfectly similar and -1 means they are perfectly dissimilar (i.e., mirror opposites).

Upon examining Table \ref{tab:alg-comparison}, the N-BEATS model stands out with the best prediction performance on three key metrics: RE, MAE, and CS. Interestingly, RandomForest performs best in terms of DTW. Deep learning methods encountered challenges in prediction due to the limited training data (only three days). The DMD-based methods, HDMD and TrHDMD, ranked consistently in the top four across all metrics. Notably, TrHDMD with $h=250$ nearly matches N-BEATS in CS, indicating its ability to capture temporal traffic trends accurately. TrHDMD also shows modest improvement over HDMD in all metrics with transferred period information, with the exception of a tie in CS at $h=300$. 

\subsection{A closer look at transfer}\label{sec:exp-transfer-beyond}
We first underscore the significant advantage of our proposed TrHDMD method, which exhibits minimal dependency on parameter tuning. In Table \ref{tab:numOfPara}, we provide a comparison of the parameters required by algorithms that have demonstrated competitive performance in previous experiments. It's worth noting that we only list a subset of the parameters from machine learning models that necessitate tweaking during training and significantly impact results. By no means is this an exhaustive list of all tunable parameters. For instance, the random forest model from the sklearn library involves approximately 19 tunable parameters, while XGBoost requires the configuration of various parameters depending on the choice of boosters and learning scenarios. As depicted in Table \ref{tab:numOfPara}, modern machine learning methods like XGBoost and N-BEATS typically involve numerous parameters compared to traditional statistical methods like VARIMA. In contrast, the TrHDMD framework relies solely on two parameters: the delay ($h$) and the threshold for shared eigenvalues ($\epsilon$). The selection of $h$ can be easily accomplished by a grid search on a bounded interval and by examining the sinusoidal shape of the right singular vectors derived from the tall Hankelized discrete time snapshot matrix. Meanwhile, $\epsilon$ determines the level of conservatism in identifying numerically close periods as identical periods. This minimal dependency on parameter tuning not only simplifies the implementation process but also enhances the framework's efficiency and effectiveness in analyzing cross-city traffic dynamics.

\begin{table}[!ht]
\caption{TrHDMD does not
rely heavily on precise parameter-tuning and only has two (hyper-)parameters (delay $h$ and the threshold for shared eigenvalues $\epsilon$).\label{tab:numOfPara}}
\centering
\begin{tabular}{|l | p{2cm} | p{4cm} |}
\hline
Method & \#Main Parameters  & Interpretation\\
\hline
VARIMA & 3  & AR Order; MA window size; Order of Differentiation\\
& & \\
RandomForest & $\gg$4  & Number of decision trees, Max depth, Max features, Maximum number of samples used in bootstrap, ...\\
& & \\
XGBoost & depends on learning scenario & Step size shrinkage; Minimum loss reduction; Dropout rate; Regularization terms, Subsample ratio of the training instances, ... \\
& & \\
N-BEATS & depends on architecture & Number of blocks/stacks/layers, Layer widths, dropout probability, Activation function, Batch size, Number of epochs, ... \\
& & \\
HDMD & 1 & Delay \\
& & \\
TrHDMD & 2 & Delay, Threshold for shared eigenvalues \\
\hline
\end{tabular}
\end{table}

\begin{figure*}[!ht]
	\centering
	\includegraphics[width=0.98\textwidth]{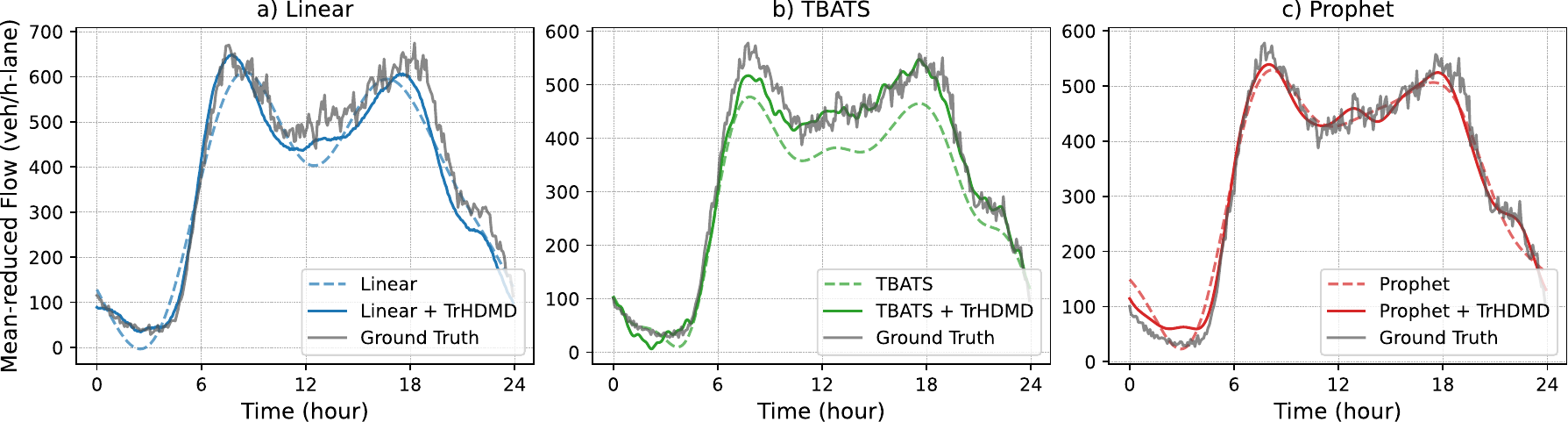}
	\caption{Linear Regression, TBATS, and Prophet algorithm enhanced by shared periods identified by TrHDMD: Linear Regression, TBATS (with naive guess) and Prophet are denoted with dashed lines. The corresponding TrHDMD-enhanced results are denoted with continuous lines. We can observe that (a) All compared algorithms have small training errors. (b) Substantial performance improvements are achieved when incorporating shared periods as custom seasonalities.}
	\label{fig:TBATS}
\end{figure*}

In our final analysis, we evaluate how the cycle times identified through TrHDMD can enhance time series forecasting methods that benefit from prior knowledge of data seasonality. We consider two algorithms for this experiment: TBATS (Trigonometric seasonality, Box-Cox transformation, ARMA errors, trend and seasonal components) \citep{de2011forecasting} and Prophet \citep{taylor2018forecasting}. These algorithms are chosen because they both can utilize custom seasonalities and are optimized when provided with appropriate seasonal patterns. Moreover, they have demonstrated success in various applications \citep{khayyat2021time,toharudin2023employing,karabiber2019electricity,abotaleb2022state}. Typically, seasonalities such as daily, weekly, monthly, and yearly are assumed based on simple heuristics due to the scarcity of comprehensive data. However, our framework offers a novel approach to enriching our understanding of data seasonality beyond these heuristics by uncovering latent patterns that can serve as valuable prior information.

We explored this concept by enhancing the performance of Linear Regression, TBATS, and Prophet with TrHDMD. Linear Regression (LR) is trained with pairs of Fourier terms $\sin{(\frac{2\pi kt}{T})}$ and $\cos{(\frac{2\pi kt}{T})}$ as independent variables, where $k$ is the higher-order harmonics and $T$ is a given period. TBATS is designed to model complex seasonal patterns through exponential smoothing for univariate time series. Prophet is an open-source software released by Facebook, it forecasts univariate time series using an additive model to fit non-linear trends with different seasonal and holiday effects. For our purpose, we select an average reading from 30 randomly chosen detectors in Stuttgart as the univariate time series. Our experiments compare the following scenarios\footnote{In this experiment, we run TBATS omitting the Box-Cox transformation and ARMA error modeling. We tested TBATS with its default configuration (including the omitted elements) and observed diminished prediction results.}: 
\begin{itemize}
    \item LR: Linear Regression with Fourier terms as independent variables, $T$ set to 24-hour (naive heuristics), $k=3$
    \item LR + TrHDMD: Linear Regression with Fourier terms as independent variables, $T$ set to custom seasonalities set as shared periods found by TrHDMD, $k=3$
    \item TBATS + TrHDMD: TBATS with custom seasonalities set as shared periods found by TrHDMD
    \item Prophet: Prophet in its default settings
    \item Prophet + TrHDMD: Prophet with custom seasonalities set as shared periods found by TrHDMD
    \item True: The ground truth time series
\end{itemize}
 The results are depicted in Fig.~\ref{fig:TBATS}, and the mean square error results are summarized in Table \ref{tab:enhance-by-TrHDMD}. We observe that for both TBATS and Prophet, the in-sample and out-of-sample errors drop after incorporating TrHDMD-derived periods ($h=300$). For In-sample errors, the percentage improvement is 73.34\% for Linear Regression, 15.19\% for TBATS and 65.13\% for Prophet. For out-of-sample errors, the percentage improvement is 28.76\% for Linear Regression, 82.84\% for TBATS, and 44.46\% for Prophet. This suggests that by introducing shared periods, not only do both forecast models improve in capturing the underlying patterns and fluctuations in the training data, but they also generalize well to new data. The fact that both errors decreased confirms that the seasonal patterns identified by TrHDMD from the source cities are relevant to the traffic dynamics of the target city. This outcome underscores our framework's value in providing rich prior information and improving forecasting accuracy, especially in data-sparse, zero-shot forecasting situations.

 \begin{table}[!ht]
\caption{Time series forecasting algorithms with custom seasonalities are enhanced by shared periods identified by TrHDMD\label{tab:enhance-by-TrHDMD}}
\centering
\begin{tabular}{|l|l|l|}
\hline
MSE & LR & LR + TrHDMD\\
\hline
In-sample & 1915.874  & 510.814\\
\hline
Out-of-Sample & 2702.550  & 1925.302\\
\hline
\hline
MSE & TBATS & TBATS + TrHDMD\\
\hline
In-sample & 215.711  & 182.942\\
\hline
Out-of-Sample & 3557.364  & 610.445\\
\hline
\hline
MSE & Prophet & Prophet + TrHDMD\\
\hline
In-sample &678.650 & 236.619\\
\hline
Out-of-Sample & 909.673 & 505.272\\
\hline
\end{tabular}
\end{table}

\begin{figure*}[!]
\centering
\includegraphics[width=1\textwidth]{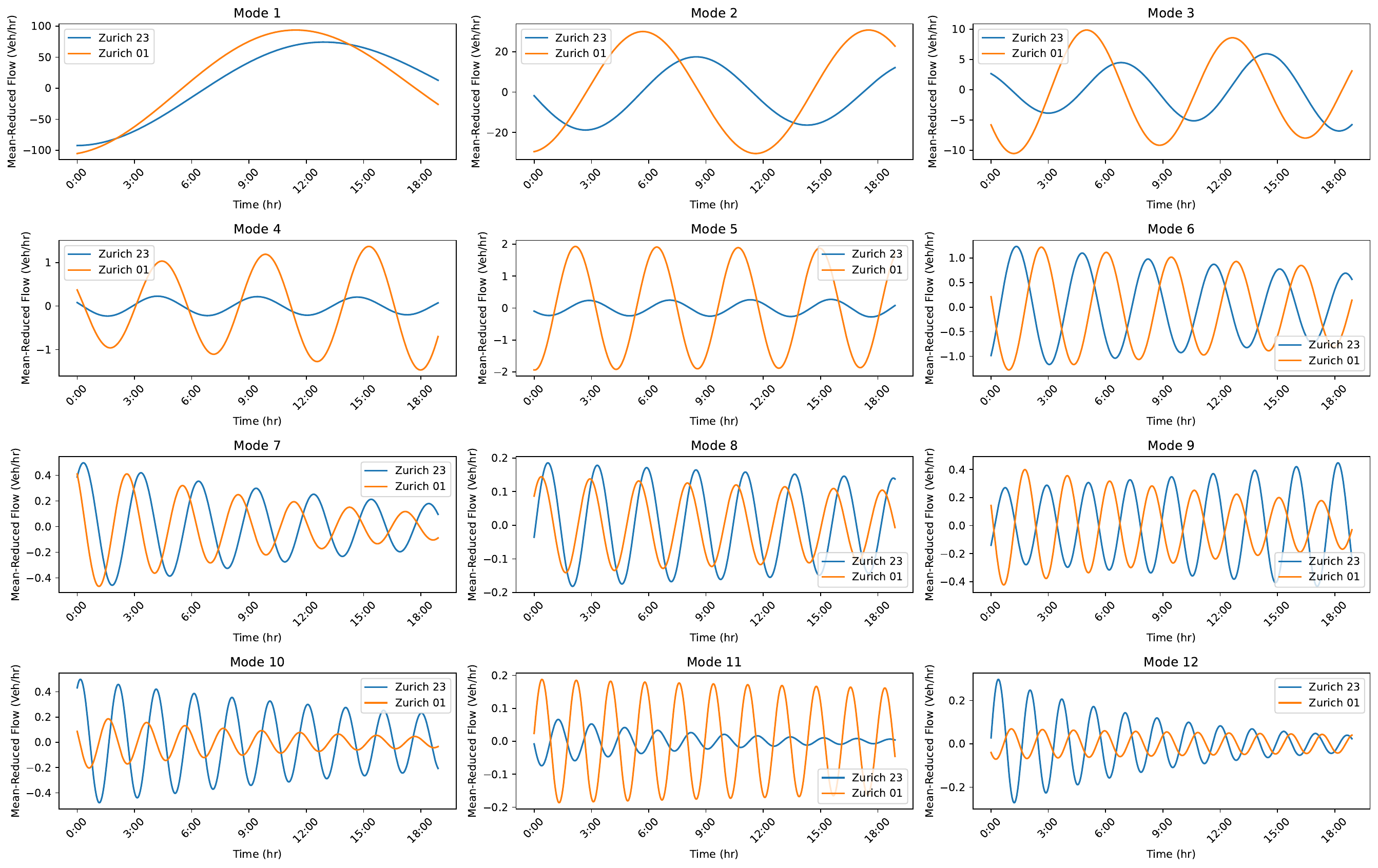}

\caption{Comparison of the top 12 subpatterns in Zurich, Switzerland, observed on September 1, 2017, and September 23, 2017. Data is aggregated over a 3-minute interval. The subpatterns are averaged over modes of 45 randomly sampled detectors, computed with delay $h= 100$. We observe that the frequencies (periods) corresponding to each pair of compared modes are very similar. However, for September 23, the top modes exhibit lower amplitudes, and the peak of the wave patterns occurs later in these dominant modes. These observations suggest lower and delayed peak flows in the Macroscopic Fundamental Diagram (MFD), consistent with hysteresis effects.}
\label{fig:zurich-comparison}
\end{figure*}

\subsection{Change of heartbeats}\label{sec:new-phenomenon}
In addition to its ability to identify transferable patterns and enhance prediction in data-scarce cities, another advantage of employing Koopman methods in our framework is the ability to detect and validate changes in traffic patterns within a single city. As an additional experiment, we examined detector readings on two different dates (September 1, 2017, and September 23, 2017) from Zurich, Switzerland. \citet{ambuhl2021disentangling} demonstrated that the dynamic Macroscopic Fundamental Diagram (dMFD) exhibited hysteresis loops on September 23, distinct from its behavior on September 1 (see Fig.~2a in \citep{ambuhl2021disentangling}). Similarly, we analyzed the city heartbeats on these dates by comparing their top 12 modes in Fig.~\ref{fig:zurich-comparison}. We observed that the frequencies (periods) of each pair of compared modes were very similar (disregarding decay rates), indicating that the analyzed traffic dynamics were generally consistent, which is expected as both datasets are from the same city. Furthermore, we noted that the top modes associated with September 23 exhibited lower amplitudes, and the peaks of the wave patterns occurred later in the dominant modes 1-3. These findings align with the hysteresis effects illustrated in Fig.~4 of \citep{ambuhl2021disentangling}, where the average flow during unloading phases was slightly lower than during loading phases, resulting in lower and delayed peak flows due to hysteresis effects affecting the flow-time relationship.

\section{Conclusion}\label{sec:conclusion}
We propose Koopman methods to identify distinct and synchronized spectral characteristics among cities with abundant traffic data, which we term ‘city heartbeats.’ Through the examination of shared cross-city patterns from source cities and their transfer as prior knowledge to the target city, we demonstrate the feasibility of knowledge sharing across diverse urban settings. Our TrHDMD framework exhibits several key advantages, including minimal dependency on parameter tuning, invariant measurements across different cities, and interpretable shared features represented by characteristic cycle times/frequencies of the common modes across cities. TrHDMD provides a framework to look into the unique schedule and universal traffic patterns of traffic dynamics at a city scale. Through extensive experiments on real-world traffic datasets. We also validate the effectiveness of our approach in traffic flow prediction tasks, both within and beyond the Koopman framework. This validation underscores its potential to contribute to transport management, particularly in cities with limited physical instrumentation. 

We focus specifically on the spectral characteristics of city traffic rather than the multi-city spatiotemporal dynamics. This choice aligns with our goal of transferring information to enhance data-scarce cities using insights from data-rich cities. Incorporating spatial information from the source city can introduce complexities for several reasons: (1) Spatio-temporal correlations vary significantly across different cities. (2) Capturing accurate long-term spatio-temporal patterns in a target city with limited data remains challenging. (3) Even in data-rich cities, issues arise from uneven spatial data distribution. The literature review shows that the spatio-temporal characteristics in cross-city knowledge transfer within smart city frameworks are under-explored. Notably, \citep{yao2019learning} is one relevant study on multi-city spatio-temporal pattern transfer and prediction, proposing a global memory framework that learns from multiple source cities to enhance spatial-temporal predictions in target cities. However, the black-box nature of this meta-learning approach leads to identified spatial-temporal patterns encrypted and encoded for the defined neural network architecture only, essentially lacking insights into spatial correlations, regional functionalities, or traffic phenomenon interpretability.

We also acknowledge the limitations of this work: Due to the lack of data on consecutive dates, our experimentation with TrHDMD is confined to training with only three days of data. Consequently, the transferred patterns capture short cycle times within a day, leaving longer seasonal and cross-city patterns unexplored. Additionally, we lose information about dynamics in the spatial dimension with input formatting when we sample city loop detectors. Future research endeavors should focus on investigating and discovering these longer patterns using consecutive weekly or monthly multi-city data. Additionally, as the sizes of the datasets increase, computation issues may arise as the Koopman methods require performing SVDs on large matrices, particularly with size increases caused by longer durations or larger numbers of detectors. Addressing these challenges necessitates exploring more efficient matrix decomposition algorithms to expedite the process and study how such approximations influence the extracted patterns.

Moreover, while we briefly mentioned the inspiring generative algorithms from Zero/Few-Shot Learning in traffic prediction, we observe that generative algorithms like Timeseries DGAN are strong in generating synthetic time series of the same period as the training time series but struggle to generate predictions for given time series (see in Table \ref{tab:alg-comparison}).  Exploring the integration of the extracted periods from our framework to enhance the performance of these generative algorithms presents an intriguing avenue for future investigation. This endeavor holds the potential for improving the predictive capabilities of such algorithms and warrants further exploration in subsequent studies.

\section*{Acknowledgement}
This work was supported in part by the NYUAD Center for Interacting Urban Networks (CITIES), funded by Tamkeen under the NYUAD Research Institute Award CG001, and in part by the NYUAD Research Center on Stability, Instability, and Turbulence (SITE), funded by Tamkeen under the NYUAD Research Institute Award CG002. The views expressed in this article are those of the authors and do not reflect the opinions of CITIES, SITE, or their funding agencies.

\bibliographystyle{apsrev4-2}
\bibliography{sample}
	
\end{document}